\documentclass[11pt,a4paper]{article}
\pdfoutput=1
\usepackage{jheppub}

\usepackage{setspace}
\usepackage[toc,page]{appendix}
\usepackage{longtable}
\usepackage{answers}
\usepackage{array}
\usepackage{enumitem}
\usepackage{graphicx}
\usepackage{color}
\usepackage{dsfont}
\usepackage{hyperref} 
\hypersetup{
    colorlinks=true, 
    linktoc=all,     
    linkcolor=black,  
    citecolor=blue,
    filecolor=blue,
    urlcolor=blue
}
\usepackage{amsmath}

\usepackage{tikz}
\usetikzlibrary{arrows,shapes}
\usetikzlibrary{trees}
\usetikzlibrary{patterns}
\usetikzlibrary{snakes}
\usetikzlibrary{matrix,arrows} 				
\usetikzlibrary{positioning}				
\usetikzlibrary{calc,through}				
\usetikzlibrary{decorations.pathreplacing}  
\usepackage{pgffor}							

\usetikzlibrary{decorations.pathmorphing}	
\usetikzlibrary{decorations.markings}
\tikzset{
    vector/.style={decorate, decoration={snake}, draw},
	provector/.style={decorate, decoration={snake,amplitude=2.5pt}, draw},
	antivector/.style={decorate, decoration={snake,amplitude=-2.5pt}, draw},
        smallvector/.style={decorate, decoration={snake,amplitude=1.5pt,post length=0.5mm}, draw},
    fermion/.style={draw=black, postaction={decorate},
        decoration={markings,mark=at position .55 with {\arrow[draw=black]{>}}}},
    fermionbar/.style={draw=black, postaction={decorate},
        decoration={markings,mark=at position .55 with {\arrow[draw=black]{<}}}},
    fermionnoarrow/.style={draw=black},
    gluon/.style={decorate, draw=black,
        decoration={coil,amplitude=4pt, segment length=5pt}},
    scalar/.style={dashed,draw=black, postaction={decorate},
        decoration={markings,mark=at position .55 with {\arrow[draw=black]{>}}}},
    scalarbar/.style={dashed,draw=black, postaction={decorate},
        decoration={markings,mark=at position .55 with {\arrow[draw=black]{<}}}},
    scalarnoarrow/.style={dashed,draw=black},
    electron/.style={draw=black, postaction={decorate},
        decoration={markings,mark=at position .55 with {\arrow[draw=black]{>}}}},
    bigvector/.style={decorate, decoration={snake,amplitude=4pt}, draw},
    arrow/.style={draw=black, postaction={decorate},
        decoration={markings,mark=at position 1 with {\arrow[draw=black]{>}}}},
}

\tikzstyle{block} = [draw, rectangle, 
    minimum height=3em, minimum width=6earticlem]

\begin{document}

\preprint{LCTP-20-01}

\title{The Black Hole Weak Gravity Conjecture with Multiple Charges}

\author[a]{Callum R. T. Jones}
\author[a]{Brian McPeak}

\affiliation[a]{Leinweber Center for Theoretical Physics, Randall Laboratory of Physics, The University of Michigan, Ann Arbor, MI 48109-1040}


\emailAdd{jonescal@umich.edu}
\emailAdd{bmcpeak@umich.edu}

\abstract{
We study the effect of higher-derivative corrections on asymptotically flat, four-dimensional, dyonic black holes in low-energy models of gravity coupled to $N$ $U(1)$ gauge fields. For large extremal black holes, the leading $\mathcal{O}\left(1/Q^2\right)$ correction to the extremality bound is calculated from the most general low-energy effective action containing operators with up to four derivatives. Motivated by the multi-charge generalization of the Weak Gravity Conjecture, we analyze the necessary kinematic conditions for an asymptotically large extremal black hole to decay into a multi-particle state of extremal black holes. In the large black hole regime, we show that the convex hull condition degenerates to the requirement that a certain quartic form constructed from the Wilson coefficients of the four-derivative effective operators, is everywhere positive. Using on-shell unitarity methods, we show that higher-derivative operators are renormalized at one-loop only if they generate local, on-shell matrix elements that are invariant tensors of the electromagnetic duality group $U(N)$. The one-loop logarithmic running of the four-derivative Wilson coefficients is calculated and shown to imply the positivity of the extremality form at some finite value of $Q^2$. This result generalizes an argument recently given by Charles \cite{Charles:2019qqt}, and shows that under the given assumptions the multi-charge Weak Gravity Conjecture is not a Swampland criterion. }

\arxivnumber{1908.10452}

\maketitle
\flushbottom

\section{Introduction}
\label{sec:intro}

String theory is widely believed to provide a UV complete description of quantum gravity. The theory is also believed to admit an astronomical number of vacua, which manifest at low energies as effective field theories (EFTs). This set of consistent string vacua is known as the \textit{Landscape}. Due to the large number of low-energy descriptions, it may be difficult or impossible to find a vacuum that describes our world. Recently a different approach has proven useful: rather than searching through vacua, we should study the general conditions under which an EFT admits a UV completion that includes quantum gravity. Theories that admit no such completion are said to be in the \textit{Swampland} \cite{Vafa:2005ui}. A number of Swampland criteria have been put forward (for a review of the program, see \cite{Brennan:2017rbf, Palti:2019pca}). In practice, Swampland criteria are proposed and supported using very different approaches. One is to study features present in known compactifications of string theory. Another approach relies on determining which properties of the infrared are required for consistency, and then studying how this constrains physics in the ultraviolet. Both sources are indirect, which makes rigorous proofs of the Swampland conjectures elusive.

One compelling candidate for a general principle constraining consistent string vacua is the \textit{weak gravity conjecture} (WGC) \cite{ArkaniHamed:2006dz}. Various forms of the conjecture have been proposed, but roughly it states that EFTs that arise as low energy descriptions of theories of quantum gravity must have a state with a greater charge than mass-- i.e. for which ``gravity is the weakest" force. Were this not the case, extremal or near-extremal black holes would unable to decay because emitting a sub-extremal state would cause the left-over black hole to be superextremal, violating cosmic censorship. This, in turn is problematic because it leads to the existence of an arbitrarily large number of stable states, which is believed to be pathological \cite{Vafa:2005ui}. We now review these arguments in more detail.

\subsection{Review of the Weak Gravity Conjecture}
\label{subsec:review}

 The original Weak Gravity Conjecture (WGC) was formulated as a Swampland criterion \cite{ArkaniHamed:2006dz}: \textit{in a UV complete model of quantum gravity, there should not exist an infinite tower of exactly stable states in a fixed direction in charge space.} Arguments against such an infinite tower include that it might lead to a species problem or remnant issues \cite{Susskind:1995da, Banks:2006mm}. No proof of this statement has been given, but it is consistent with all known explicit examples of string compactifications and is conceptually consistent with a number of other conjectures about quantum gravity, such as the finiteness principle and the absence of global symmetries \cite{Vafa:2005ui}.

The conjecture can be equivalently interpreted as a statement about the (in-)stability of asymptotically large extremal black holes. In quantum gravity, \textit{elementary} states with super-Planckian masses can be expected to appear to distant observers as black hole solutions of some low-energy effective field theory (EFT) \cite{Susskind:1993ws,Horowitz:1996nw}. The decay of such a state must have an equivalent semi-classical description as the discharge of the black hole, for example by Schwinger pair production of charged states near the horizon \cite{Gibbons:1975kk}. Since the relevant energy scale $\mu$ for the EFT calculation is here given by the scale of the black hole horizon $\mu \sim M_{\text{Pl}}^2/M$, asymptotically large black holes are well approximated by standard two-derivative Einstein gravity together with any additional massless degrees of freedom. All other details of the UV physics are integrated out and appear in the low-energy EFT as contributions to Wilson coefficients of higher-derivative effective operators that give subleading corrections to the black hole solutions. Models of quantum gravity can then be organized into universality classes according to their massless spectra and lowest dimension interactions; each class of model has an associated set of large black hole solutions that must then correspond to the asymptotic spectrum of super-Planckian elementary states. 

In this paper we consider the universality class of models in four-dimensions with zero cosmological constant and a massless spectrum of matter fields consisting of $N$ $U(1)$ gauge fields. To begin we review the statement of the WGC for $N=1$; in this class the spectrum of large black holes corresponds to the familiar Kerr-Newman solutions. Within a given charge sector, the lightest black hole corresponds to the extremal, non-rotating solution with $Q^2=M^2/M_{\text{Pl}}^2$. If the WGC is true, then for all $Q^2$ greater than some critical value, the corresponding extremal black hole must be able discharge. Whether this is kinematically possible depends on the spectrum of charged states with masses lighter than the black hole. For a general transition of the form
\begin{equation}
    |Q,M\rangle \rightarrow |q_1,m_1\rangle \otimes |q_2,m_2\rangle \otimes ... \otimes|q_n,m_n\rangle,
\end{equation}
where each of the final states is assumed to be localized and at rest asymptotically far away (with zero kinetic and gravitational potential energy), conservation of total energy and total charge requires
\begin{equation}
    Q = q_1+q_2+...+q_n, \hspace{10mm} M = m_1+m_2+...+m_n.
\end{equation}
If the initial state is a large extremal black hole with $Q^2 = M^2/M_{\text{Pl}}^2$, then at least one of the daughter states $|q_i,m_i\rangle$ must be \textit{self-repulsive}, meaning $q_i^2 \geq m_i^2/M_{\text{Pl}}^2$ (regardless of whether we include higher-derivative corrections). Conversely, if there are no self-repulsive states then such a decay is impossible and an infinite tower of extremal black holes are exactly stable, violating the aforementioned Swampland criterion. This leads to the common formulation of the WGC:\\

\textbf{Weak Gravity Conjecture (Single Charge):} \textit{In a UV complete model of quantum gravity there must exist some state with} $Q^2 \geq M^2/M_{\text{Pl}}^2$.\\

In the context of a specific model, to show that the WGC is violated requires complete knowledge of the spectrum of charged states. To show that it is satisfied however, requires only the existence of a single self-repulsive state. It is useful to separate charged states into three regimes according to their masses:

\begin{enumerate}
	\item \textbf{Particle regime} ($M\ll M_{\text{Pl}}$): States in this regime are well-described by ordinary quantum field theory on a fixed spacetime background. 
	\item \textbf{Stringy regime} ($M \lesssim M_{\text{Pl}}$): States in this regime are intrinsically related to the UV completion. They can usually only be calculated from a detailed understanding of the UV physics such as an explicit string compactification.
	\item \textbf{Black hole regime} ($M \gg M_{\text{Pl}}$): States in this regime are well-described by classical black hole solutions in the relevant low-energy model of gravity. 
\end{enumerate}

In this paper we are analyzing the spectrum of charged states in the black hole regime. The corresponding analysis for a single $U(1)$ gauge field was made in \cite{Kats:2006xp}; we begin by reviewing their discussion. Naively, it would seem impossible for a charged black hole to be self-repulsive since this would violate the extremality bound. The usual bound $Q^2\leq M^2/M_{\text{Pl}}^2$ is derived by requiring the existence of a horizon (by requiring Weak Cosmic Censorship). When the higher derivative corrections to the effective action are included, the black hole solutions and the associated extremality bounds are modified. For large black holes, with $Q^2\gg 1$, these corrections can be calculated perturbatively in $1/Q^2$, with the leading corrections corresponding to four-derivative effective operators. The authors of \cite{Kats:2006xp} analyzed electrically charged solutions to the following effective action
\begin{equation}
\label{4dersingle}
S = \int \text{d}^4 x \sqrt{-g}\left[\frac{M_{\text{Pl}}^2}{4}R - \frac{1}{4}F_{\mu\nu}F^{\mu\nu} + \alpha \left(F_{\mu\nu} F^{\mu\nu}\right)^2  + \beta  \left(F_{\mu\nu} \tilde{F}^{\mu\nu}\right)^2 +\gamma F_{\mu\nu}F_{\rho\sigma}W^{\mu\nu\rho\sigma}\right] \, ,
\end{equation}
where $W^{\mu \nu \rho \sigma}$ is the Weyl tensor. To leading-order, the corrected extremality bound is 
\begin{equation}
    \frac{M_{\text{Pl}}^2Q^2}{M^2} \leq 1+\frac{4}{5Q^2}(2\alpha -\gamma)+\mathcal{O}\left(\frac{1}{Q^4}\right).
\end{equation}
The $\mathcal{O}\left(1/Q^4\right)$ contributions correspond to next-to-leading-order in the four-derivative operators and leading-order in six-derivative operators. If the corrected extremality bound is positive
\begin{equation}\label{singleext}
    2\alpha-\gamma>0,
\end{equation}
then extremal black holes with finite charge are self-repulsive and the WGC is satisfied in the black hole regime. Conversely, if the corrected extremality bound is negative
\begin{equation}
    2\alpha-\gamma<0,
\end{equation}
then the decay of asymptotically large extremal black holes into extremal black holes with large but finite charge is kinematically impossible. This does not mean that the WGC is violated, but rather that if it is valid then there must exist a self-repulsive state in either the stringy or particle regimes. 

Various arguments have been given that (\ref{singleext}) should always be true, even from a low-energy perspective. These include arguments from unitarity, causality \cite{Hamada:2018dde}, positivity of the S-matrix \cite{Bellazzini:2019xts}, shifts to entropy bounds \cite{Cheung:2018cwt}, and renormalization group running \cite{Charles:2019qqt}. \\
\\
The purpose of this paper is to generalize the above discussion to the universality class of models for which the low-energy matter spectrum consists of $N$ $U(1)$ gauge fields. We consider black hole solutions with general electric and magnetic charges. 

The two-derivative approximation to the EFT has many accidental symmetries, including an $O(N)$ global flavor symmetry, parity and $U(N)$ electromagnetic duality symmetry. We do not assume that any of these symmetries are preserved in the UV, and instead analyze the most general possible EFT with the assumed low-energy spectrum
\begin{align}
  \begin{split}
    S &= \int \mathrm{d}^4x \sqrt{- g} \Big[ \frac{M_{\text{Pl}}^2}{4}R - \frac{1}{4} F^i_{\mu \nu} F^{i \, \mu \nu} 
    + a_{ijk}F^i_{\mu\nu}F^{j\nu\rho}{F^k_{\rho}}^\mu+ b_{ijk}F^i_{\mu\nu}F^{j\nu\rho}{\tilde{F}^k_{\rho}} {}^\mu \\
    &\hspace{25mm}+  \alpha_{i j k l} \, F^i_{\mu \nu} F^{j \, \mu \nu} F^k_{\rho \sigma} F^{l \, \rho \sigma}
    + \beta_{i j k l} \,  F^i_{\mu \nu} \tilde{F}^{j \, \mu \nu} F^k_{\rho \sigma} \tilde{F}^{l \, \rho \sigma} 
    \\
    & \qquad \qquad \qquad + \gamma_{i j} \, F^i_{\mu \nu} F^j_{\sigma \rho} W^{\mu \nu \sigma \rho}
    + \chi_{i j k l} \, \tilde{F}^i_{\mu \nu} F^{j \, \mu \nu} F^k_{\rho \sigma} F^{l \, \rho \sigma} 
    + \omega_{i j} \, F^i_{\mu \nu} \tilde{F}^j_{\sigma \rho} W^{\mu \nu \sigma \rho} \Big].
  \end{split}
  \label{Action}
\end{align}

In \cite{Cheung:2014vva} it was shown that the kinematic condition for a large extremal black hole with multiple charges to decay is a non-trivial generalization of the single charge version of the WGC. In general, if a set of \textit{light} states $|\vec{q}_i,m_i\rangle$ are available with masses $m_i$ and charge vectors $\vec{q}_i$, then the possible charge-to-mass ratio vectors of the associated multi-particle states $|\vec{q}_1,m_1\rangle^{\otimes N_1}\otimes |\vec{q}_2,m_2\rangle^{\otimes N_2}\otimes ...$ are given by

\begin{equation}
    \vec{z} \in \left\{\frac{M_{\text{Pl}}\sum_i N_i \vec{q}_i}{\sum_i |N_i| m_i}, \;\; N_i \in \mathds{Z}\right\}.
\end{equation}
Here $N_i<0$ corresponds to contributions from CP conjugate states. This set describes the convex hull of the charge-to-mass vectors $\vec{z}_i = \vec{q}_i/m_i$. The condition that the decay of asymptotically large extremal black holes be allowed is given by the \textit{convex hull condition}\cite{Cheung:2014vva}:\\

\textbf{Weak Gravity Conjecture (Multiple Charges):} \textit{In a UV complete model of quantum gravity, the convex hull of the set of charge-to-mass vectors}
\begin{equation}
    \vec{z}_i \equiv \frac{M_{\text{Pl}}}{m_i}
        \begin{pmatrix}
            \vec{q}_i\\
            \vec{p}_i
        \end{pmatrix},
\end{equation}
 \textit{for every charged state in the spectrum, with mass $m$, electric charges $\vec{q}=(q^1,q^2...)$ and magnetic charges $\vec{p}=(p^1,p^2,...)$, must enclose the unit ball $|\vec{z}|^2\leq 1$.}\\

As in the single charge case, to show that a given model does not satisfy this condition requires complete knowledge of the spectrum of charged states. It is however possible to show that this condition is satisfied with only partial knowledge of the spectrum since the convex hull of a subset of vectors always forms a subregion of the full convex hull. This condition has been previously analyzed from several perspectives \cite{Andriolo:2018lvp}, considering contributions from the \textit{particle regime}. In this paper, we will describe the general conditions on the Wilson coefficients $\{a_{ijk},b_{ijk},\alpha_{ijkl},\beta_{ijkl},\gamma_{ij},\chi_{ijkl},\omega_{ij}\}$ under which the convex hull condition is satisfied by contributions from the \textit{black hole regime}. 

\subsection{Overview of Results}
\label{subsec:results}

This paper is organized as follows. In section \ref{sec:extshift}, we calculate the leading-order corrections to dyonic, non-rotating, extremal black hole solutions corresponding to the effective action (\ref{Action}); various technical details are given in appendices \ref{appMax} and \ref{appLag}. The corrected extremality bound is inferred by demanding the existence of a horizon (\ref{extremality result}) and is found to depend on all five of the four-derivative operators, including parity violating operators when magnetic charges are present. It is shown that the three-derivative operators do not give corrections to spherically symmetric solutions at any order in the perturbative expansion. 

In section III, we describe the necessary kinematic conditions for asymptotically large black holes to decay into finite charge black holes. First we describe the natural generalization of the convex hull condition to the black hole regime, then we argue (with a proof relegated to appendix \ref{appConvex}) that in the large black hole regime, when the perturbative expansion in $1/Q^2$ is justified, the extremality surface is always convex. The black hole WGC is then shown to reduce to the condition that a quartic form (\ref{extremalityform}) is everywhere positive. We comment on the implications of known unitarity and causality constraints on the Wilson coefficients. The condition is analyzed in detail in two illustrative examples; first we consider the black hole that is charged under two electric charges $q_1$ and $q_2$, and second we consider the black hole that has both an electric charge $q$ and a magnetic charge $p$ under a single $U(1)$ gauge field.

In section IV we analyze the one-loop logarithmic running of the Wilson coefficients of the four-derivative effective operators. Using on-shell unitarity methods we prove that higher-derivative operators are renormalized only if they generate local, on-shell matrix elements that are invariant tensors of the maximal compact electromagnetic duality group $U(N)$. Using this non-renormalization theorem, together with the explicit one-loop UV divergence of Einstein-Maxwell theory, the logarithmic running of the Wilson coefficients is calculated and shown to imply the positivity of the extremality form (\ref{extremalityform}) at some finite charge. 

In appendix \ref{appA} we review the correspondence between non-redundant EFT operator bases and local on-shell matrix elements. Using elementary spinor-helicity methods a complete and independent basis of matrix elements is determined and the corresponding three- and four-derivative local operators constructed.

\section{Extremality Shift}
\label{sec:extshift}

In this section we will determine the effect of higher-derivative operators on the extremality bound using the method developed in \cite{Kats:2006xp}. In the case of multiple charges, this amounts to delineating the space of allowed charge combinations $Q = \sqrt{ q_1^2 + p_1^2 + ...}$ for a given mass $m$. We use the presence of a naked singularity, or absence of an event horizon, to rule out charge configurations at a given mass; such combinations of charge and mass will be called \textit{superextremal}. 

In pure Einstein-Maxwell theory, the superextremal black holes have $Q/m > 1$. We refer to such an inequality as the \textit{extremality bound}. This requirement derives from the positivity of the discriminant of the function $1/ g_{rr}$, which itself comes from the requirement that that function should have a zero (i.e. the event horizon).  We will see that the higher-derivative corrections have the effect of shifting the right-hand side of this bound by factors proportional to the Wilson coefficients and suppressed by factors of $1 / Q$. Generically, $n$-derivative operators will contribute a term in the extremality bound that is proportional to $1 / Q^{n-2}$.




This approach is necessarily first-order in the EFT coefficients; if we were to compute the shift to second-order in the four-derivative coefficients, we would need also to consider the first-order effect of six-derivative operators, as these contribute at the same order in $1/Q$. This means that at each step we eliminate all terms that are beyond leading-order in the four-derivative coefficients. 

\subsection{No Correction from Three-Derivative Operators}

When $N\geq 3$ the leading effective interactions are given by three-derivative operators:
\begin{equation}
    S_{3} = \int \text{d}^4 x \sqrt{-g}\left[\frac{M_{\text{Pl}}^2}{4}R - \frac{1}{4} F^i_{\mu \nu} F^{i \, \mu \nu} +a_{ijk}F^i_{\mu\nu}F^{j\nu\rho}{F^k_{\rho}}^\mu+ b_{ijk}F^i_{\mu\nu}F^{j\nu\rho}{\tilde{F}^k_{\rho}}{}^\mu\right],
\end{equation}
where the dual field strength tensor is defined as
\begin{equation}
    \tilde{F}^{i\mu\nu} = \frac{1}{2}\epsilon^{\mu\nu\rho\sigma}F^i_{\rho\sigma} \, .
\end{equation}
From the index structure of the three-derivative operators (alternatively from the structure of the corresponding local matrix elements given in appendix \ref{appA}) one can show that both $a_{ijk}$ and $b_{ijk}$ are totally antisymmetric. 

We analyze solutions to the equations of motion:
\begin{align}
    \label{3dereom}
    \nabla_\mu F^{i\mu\nu} &= -6a_{ijk} \nabla_\mu \left(F^{j\nu\rho}{F^k_\rho}^\mu\right) -6b_{ijk} \nabla_\mu \left({F^j_\alpha}^\nu \tilde{F}^{k\mu\alpha}\right),\nonumber\\
    R_{\mu\nu}-\frac{1}{2}Rg_{\mu\nu} &= \frac{2}{M_{\text{Pl}}^2}\left[F^i_{\mu\rho}{F^i_\nu}^\rho -\frac{1}{4}g_{\mu\nu}F^i_{\rho\sigma}F^{i\rho\sigma}\right. \nonumber\\
    & \hspace{5mm}  + 2\left. a_{ijk}\left[F^i_{\alpha\mu}F_\nu^{j\rho}F_\rho^{k\alpha}-\frac{1}{2}g_{\mu\nu}F_{\rho\sigma}^i F^{j\sigma\alpha}F_\alpha^{k\rho}\right] +2 b_{ijk}F^i_{\mu\rho}F^j_{\nu\sigma}\tilde{F}^{k\rho\sigma}\right].
\end{align}
By an elementary spurion analysis it is clear that there can be no modification of the extremality bound at $\mathcal{O}(a,b)$. Promoting $a_{ijk}$ and $b_{ijk}$ to background fields transforming as totally anti-symmetric tensors of the (explicitly broken) flavor symmetry group $O(N)$, at leading order the extremality shift can depend only on invariants of the form $a_{ijk} q^i q^j q^k$ or $a_{ijk} q^i q^j p^k$, which vanish. At next-to-leading order there could be contributions of the form $a_{ijk}a_{klm}q^ip^jq^lp^m$, which do not obviously vanish for similarly trivial reasons. If present such contributions would appear at the same order, $\mathcal{O}\left(1/Q^2\right)$ as the leading-order contributions from the four-derivative operators. 

Interestingly these $\mathcal{O}(a^2,ab,b^2)$ corrections also vanish. To show this, we evaluate the right-hand-side of (\ref{3dereom}) on a spherically symmetric ansatz, 
\begin{align}
\begin{split}
        ds^2 \ = \ & g_{tt}(r) \, dt^2 + g_{rr}(r) \, r^2 dr^2 + d \Omega^2 , \qquad F^{i \, tr}(r), \qquad   F^{i \, \theta \phi}(r), 
\end{split}
\end{align}
with the remaining components of the field strength tensors set to zero. The higher-derivative terms are seen to vanish due to the structure of the index contractions. The equations of motion for the non-zero components $g_{tt},\; g_{rr},\; F^{i t r}, \; F^{i \theta \phi}$ are \textit{identical} to the equations of motion of two-derivative Einstein-Maxwell. The Reissner--Nordstr{\"o}m black hole remains the unique spherically symmetric solution to the higher-derivative equations of motion with a given charge and mass.

It is interesting to note that the above argument fails if the solution is only axisymmetric, as in the general Kerr-Newman solution. For spinning, dyonic black holes, the three-derivative operators might give $\mathcal{O}\left(1/Q^2\right)$ corrections to the extremality bounds. We leave the analysis of this case to future work.

\subsection{Four-Derivative Operators}

The three-derivative operators have no contribution on spherically symmetric backgrounds. Thus, the leading shift to the extremality bound comes from four-derivative operators. We consider the action
\begin{align}
  \begin{split}
    S_4 &= \int \mathrm{d}^4x \sqrt{- g} \Big( \frac{R}{4} - \frac{1}{4} F^i_{\mu \nu} F^{i \, \mu \nu} 
      + \alpha_{i j k l} \, F^i_{\mu \nu} F^{j \, \mu \nu} F^k_{\rho \sigma} F^{l \, \rho \sigma}
    + \beta_{i j k l} \,  F^i_{\mu \nu} \tilde{F}^{j \, \mu \nu} F^k_{\rho \sigma} \tilde{F}^{l \, \rho \sigma} 
    \\
    & \qquad \qquad \qquad + \gamma_{i j} \, F^i_{\mu \nu} F^j_{\sigma \rho} W^{\mu \nu \sigma \rho}
    + \chi_{i j k l} \, \tilde{F}^i_{\mu \nu} F^{j \, \mu \nu} F^k_{\rho \sigma} F^{l \, \rho \sigma} 
    + \omega_{i j} \, F^i_{\mu \nu} \tilde{F}^j_{\sigma \rho} W^{\mu \nu \sigma \rho} \Big).
  \end{split}
  \label{Action4}
\end{align}
Here the Latin indices run from $1$ to the number of gauge fields $N$. This is the most general possible set of four-derivative operators for Einstein-Maxwell theory in 4 dimensions. For a thorough discussion on how these operators comprise a complete basis, see appendix \ref{appA}. We will see that the parity-odd operators can contribute if we allow for magnetic charges. Our calculation is identical to the one performed in \cite{Kats:2006xp} if we set $N \rightarrow 1$ and turn on only electric charges. We have chosen units with $ M_{\text{Pl}}= 1$ for convenience, though they may be restored via dimensional analysis. 

\subsubsection{Background}

First consider the uncorrected theory, which is gravity with $N$ $U(1)$ gauge fields. This theory admits solutions that are black holes with up to $N$ electric and magnetic charges. These solutions take the form: %
\begin{align}
\label{zerosol}
\begin{split}
        ds^2 \ = \ & g_{tt} \, dt^2 + g_{rr} \, dr^2 + r^2 d \Omega^2 , \qquad F^{i \, tr} \ = \ \frac{q^i}{r^2}, \qquad  F^{i \, \theta \phi} \ = \ \frac{p^i}{r^4 \, \sin \theta} \, , \\[5pt]
        & \qquad -g_{tt} \ = \ g^{rr} \ = \ 1 - \frac{2 M}{r} + \frac{Q^2}{r^2} \, .
\end{split}
\end{align}
Here $Q^2 = q^i q^i + p^i p^i$. These backgrounds are spherically symmetric, so we will impose this as a requirement on the shifted background\footnote{Spherical symmetry ensures that $1/g_{rr} = g^{rr}$, even for the corrected solutions. However, $g_{tt}$ and $1/g_{rr}$ will generally receive different corrections, which is why we do not denote these functions with one symbol such as $f(r)$.}. In the case of spherical symmetry, one may rearrange the Einstein equation and integrate to find \cite{Kats:2006xp}
\begin{align}
    \label{T00Int}
    g^{rr} = 1 - \frac{2 M}{r} - \frac{2}{r} \int_r^\infty dr r^2 T_t{}^t  \, .
\end{align}
For the uncorrected theory, the stress tensor is 
\begin{align}
    T_{\mu \nu} = F^i_{\mu \alpha} F^i_{\nu}{}^{\alpha} - \frac{1}{4} F^i_{\alpha \beta} F^{i \alpha \beta} g_{\mu \nu} \, .
\end{align}
In this case, it is easy to see that the effect of the stress tensor is to add the $\frac{q^2 + p^2}{r^2}$ term to $g^{rr}$.

\subsubsection{Corrections to the Background}

Now consider the effect of the four-derivative terms. To compute their effect on the geometry, we must compute their contributions to the stress tensor. We will expand the stress tensor as a power series in the Wilson coefficients as 
\begin{align}
    T = T^{(0)} + T^{(1)}_{Max} + T^{(1)}_{Lag} + ...
\end{align}
Here we have written two terms that are proportional to the first power of the Wilson coefficients $( \alpha_{ijkl}, \beta_{ijkl}, ... )$, because there are two different sources of first-order corrections.

The first change $T^{(1)}_{Max}$ comes from the effect of these operators on solutions to the Maxwell equations, which changes the values of $F^i_{\mu \alpha} F^i_{\nu}{}^{\alpha} - \frac{1}{4} F^i_{\alpha \beta} F^{i \alpha \beta} g_{\mu \nu} $. Thus, $T^{(1)}_{Max}$ essentially comes from evaluating the zeroth-order stress tensor on the first-order solution of the $F^i$ equations of motion.

The second change $T^{(1)}_{Lag}$ derives from varying the higher-derivative operators with respect to the metric. Thus, this term is essentially the first-order stress tensor, and we will evaluate it on the zeroth-order solutions to the Einstein and Maxwell equations. The remainder of this section will be devoted to computing each of these contributions. 

\subsubsection{Maxwell Corrections}

The first source of corrections to the stress tensor derives from including the corrections to the value of $F$. The corrected gauge field equation of motion is
\begin{align}
    \begin{split}
    \nabla_{\mu} F^{i \mu \nu} =& \, \nabla_{\mu} \Big( 8 \, \alpha_{ijkl}  F^{j \mu \nu} F^k_{\alpha \beta} F^{l \alpha \beta} + 8 \, \beta_{ijkl}  \tilde{F}^{j \mu \nu} F^k_{\alpha \beta} \tilde{F}^{l \alpha \beta} + 4 \, \gamma_{ij}  F^j_{\alpha \beta} W^{\mu \nu \alpha \beta}  \\
    &  \qquad \qquad + 4 \, \left( \chi_{ijkl} \tilde{F}^{j \mu \nu} F^k_{\alpha \beta} F^{l \alpha \beta} + \chi_{klij} F^{j \mu \nu} \tilde{F}^k_{\alpha \beta} F^{l \alpha \beta} \right) +  4 \, \omega_{ij}  \tilde{F}^j_{\alpha \beta} W^{\mu \nu \alpha \beta} \Big).
    \label{Maxwell}
    \end{split}
\end{align}
We denote the right-hand side of this equation by $\nabla_{\mu} G^{\mu \nu}$. The first-order solution to the Maxwell equation leads to corrections that equal (see appendix \ref{appMax})
\begin{align}
    \begin{split}
        & (T^{(1)}_{Max})_t{}^t \ = \ - \left[ \sqrt{-g} G^{i t r} \right]^{(1)} \left[  \sqrt{-g} F^{i t r} \right]^{(0)} / (g_{\theta \theta } g_{\phi \phi}) \, .
        \label{max_form}
    \end{split}
\end{align}
By plugging in the zeroth-order values of the fields into this expression, we compute the corrections to the stress tensor through the Maxwell equation:
\begin{align}
    \begin{split}
        \label{Maxwell Corrections}
        (T^{(1)}_{Max})_t{}^t  \ & = \ \frac{8}{r^8} \, \Big( 2 \alpha_{ijkl} \, q^i q^j (q^k q^l - p^k p^l) + 4 \,  \beta_{ijkl} \, q^i p^j  q^k p^l  +  2 \gamma_{ij} \, q^i q^j \, (Q^2  - Mr) \\
        & \qquad \quad + \chi_{ijkl} \, \left( q^i p^j( q^k q^l - p^k p^l)  + 2 q^i q^j q^k p^l \right) +  \, 2 \omega_{ij}\, q^i p^j \, (Q^2 - Mr)  \Big) \, .
    \end{split}
\end{align}
The details of this derivation may be found in appendix \ref{appMax}, but we should comment on a few interesting points. First, note the only $G^{i t r}$ arises in the result. This is due to the Bianchi identity, which does not allow $G^{i \theta \phi}$ to contribute. The Bianchi identity requires that $\partial_r F_{\theta \phi} = 0$, so in fact $F^i_{\theta \phi}$ can get no corrections at any order. 

A subtlety arises from the fact that the metric appears in the expression for the stress tensor. Therefore, it might appear that the first-order corrections to $T_t{}^t$ involve contributions from the first-order value of $F$ and the first-order value of $g$. This would be problematic because the first-order value of $g$ is what we use the stress tensor to compute in the first place. In fact, this is not an issue; only the zeroth-order metric shows up in (\ref{max_form}). This decoupling relies on cancellation between various factors of metric components, as well as spherical symmetry. Without this, the perturbative procedure we use to compute the shift to the metric would not work. We do not expect this decoupling between corrections to the stress tensor and corrections to the metric to happen for general backgrounds. It would be interesting to study the general circumstances under which it occurs.

\subsubsection{Lagrangian Corrections}

The second source of corrections is comparatively straightforward and comes from considering the higher-derivative terms in the Lagrangian as ``matter" and varying them with respect to the metric. The variations of each term are given in appendix \ref{appLag}. The result is 
\begin{align}
    \begin{split}
        (T^{(1)}_{Lag})_t{}^t \quad &= \quad \frac{1}{r^8} \, \Big( 4 \, \alpha_{ijkl} \, (p^i p^j p^k p^l + 2 q^i q^j p^k p^l - 3 q^i q^j q^k q^l) - 4 \, \beta_{ijkl} \, q^i p^j  q^k p^l\\
        &  -\frac{4}{3} \, \gamma_{ij} \left( q^i q^j (6Q^2 - 2 M r - 3 r^2) + p^i p^j (6Q^2 - 10 Mr - 3 r^2) \right) \\
        & - 16\,  \chi_{ijkl} \, q^i p^j q^k q^l - \frac{8}{3} \, \omega_{ij} q^i p^j(4 Mr - 3 r^2) \Big) \, .
    \end{split}
\end{align}

In both cases, we have simplified the expressions by using the symmetries of the tensor appearing in the higher-derivative terms (e.g. $\alpha_{ijkl} = \alpha_{jikl} = \alpha_{klij}$). 

\subsection{Leading Shift to Extremality Bound}

By adding together both sources of corrections and computing the integral in (\ref{T00Int}), we compute the shift to the radial function $g^{rr}$ defined as, 
\begin{align}
    g^{rr} = 1 - \frac{2 M}{r} + \frac{q^2 + p^2}{r^2} + \Delta g^{rr}.
\end{align}
Then the shift is given by
\begin{align}
    \begin{split}
        \label{metricshift}
        \Delta g^{rr} &= -\frac{4}{15 r^6} \Big( 6 \, \alpha_{ijkl} \, (q^i q^j - p^i p^j)(q^k q^l - p^k p^l) \,
         + \, 24 \beta_{ijkl} q^i p^j q^k p^l \\
        & +\gamma _{ij} \, \left( q^i q^j - p^i p^j\right) \left( 12 Q^2 -25 M r + 10 r^2 \right)  \\
        &  + 12 \,  \chi_{ijkl} \, q^i p^j \, \left(q^k q^l-p^k p^l\right) + 2 \,  \omega _{ij} \,   q^i p^j \left( 12 Q^2 -25 M r + 10 r^2 \right) \Big).
    \end{split}
\end{align}

To find the shift to extremality that results from this, we examine when the new radial function $g^{rr}(r, M, Q)$ has zeros \footnote{Equivalently we could examine the zeros of $g_{tt}$. This must give identical results since the consistency of the metric signature requires that $g_{tt}$ and $g^{rr}$ have the same set of zeros.   }. This equation is sixth order in $r$, but we are only interested in the first-order shift to the solution. We Taylor-expand near the extremal solution where $r = M$ and $Q = M$, and keep only terms that are first-order in Wilson coefficients:
\begin{align}
    \begin{split}
        g^{rr}(r, M, Q) & \, = \, g^{rr}(M, M, M) + (Q - M) \, \partial_Q g^{rr} |_{(M, M, M)} + (r - M) \,  \partial_r g^{rr} |_{(M, M, M)} \\
        & \, = \, \Delta g^{rr}(M, M, M) + (Q - M) \, \partial_Q g^{rr} |_{(M, M, M)}. \,
    \end{split}
\end{align}
We have kept $M$ fixed.  In going from the first to the second line, we have used that the uncorrected metric vanishes at $(M, M, M)$ so $g^{rr}(M, M, M) = \Delta g^{rr}(M, M, M)$. We also used that the uncorrected metric also has vanishing $r-$derivative at $(M, M, M)$, so the last term on the first line may be removed because it is second-order in Wilson coefficients. The requirement that $g^{rr}$ leads to the condition:
\begin{equation}
    g^{rr}(r, M, Q) = 0 \implies  Q - M =  - \frac{\Delta g^{rr}(M, M, M)}{ \partial_Q  g^{rr}(M, M, M)}.
\end{equation}
Now we evaluate this expression and divide by $m$ to find the result for the extremality bound $|\vec{z}|^2 = Q^2/M^2 $
\begin{align} 
    \begin{split}
        |\vec{z}| & \leq 1 +  \frac{2}{5 (Q^2)^3} \Big( 2 \, \alpha_{ijkl} \, (q^i q^j - p^i p^j)(q^k q^l - p^k p^l) \,
         + \, 8 \beta_{ijkl} q^i p^j q^k p^l - \gamma _{ij} \, \left( q^i q^j - p^i p^j\right) Q^2 \\
        &  \hspace{25mm} + 4 \,  \chi_{ijkl} \, q^i p^j \, \left(q^k q^l-p^k p^l\right) - 2 \,  \omega _{ij} \,   q^i p^j Q^2  \Big) + \mathcal{O}\left(\frac{1}{(Q^2)^2}\right).
    \label{extremality result}
    \end{split}
\end{align}
This is the main technical result of this paper. In the next section, we comment on the constraints that black hole decay might place on these coefficients, and we analyze this expression for the case of black holes with two electric charges, and the case of black holes with a single electric and single magnetic charge.

\section{Black Hole Decay and the Weak Gravity Conjecture}
\label{sec:decay}

As described by \cite{Cheung:2014vva} and reviewed in section \ref{subsec:review}, a state with charge-to-mass vector $\vec{z}$ and total charge $Q^2\equiv \sum_i((q^i)^2+(p^i)^2)$ is kinematically allowed to decay to a general multiparticle state only if $\vec{z}$ lies in the convex hull of the light charged states. In the case of asymptotically large extremal black holes decaying to finite charge black holes, the spectrum of light states corresponds to the region compatible with the extremality bound. This bound describes a surface in $z$-space of the form
\begin{equation}
    |\vec{z}|=1+T(\vec{z},Q^2),
\end{equation}
where $T\rightarrow 0$ as $Q^2\rightarrow \infty$. The convex hull condition \cite{Cheung:2014vva} has a natural generalization to the sector of extremal black hole states:\\

\textbf{Black Hole Convex Hull Condition:} \textit{It is kinematically possible for asymptotically large extremal black holes to decay into smaller finite $Q^2$ black holes only if the convex hull of the extremality surface encloses the unit ball $|\vec{z}|\leq 1$}.\\

\begin{figure}
\includegraphics[scale=0.8]{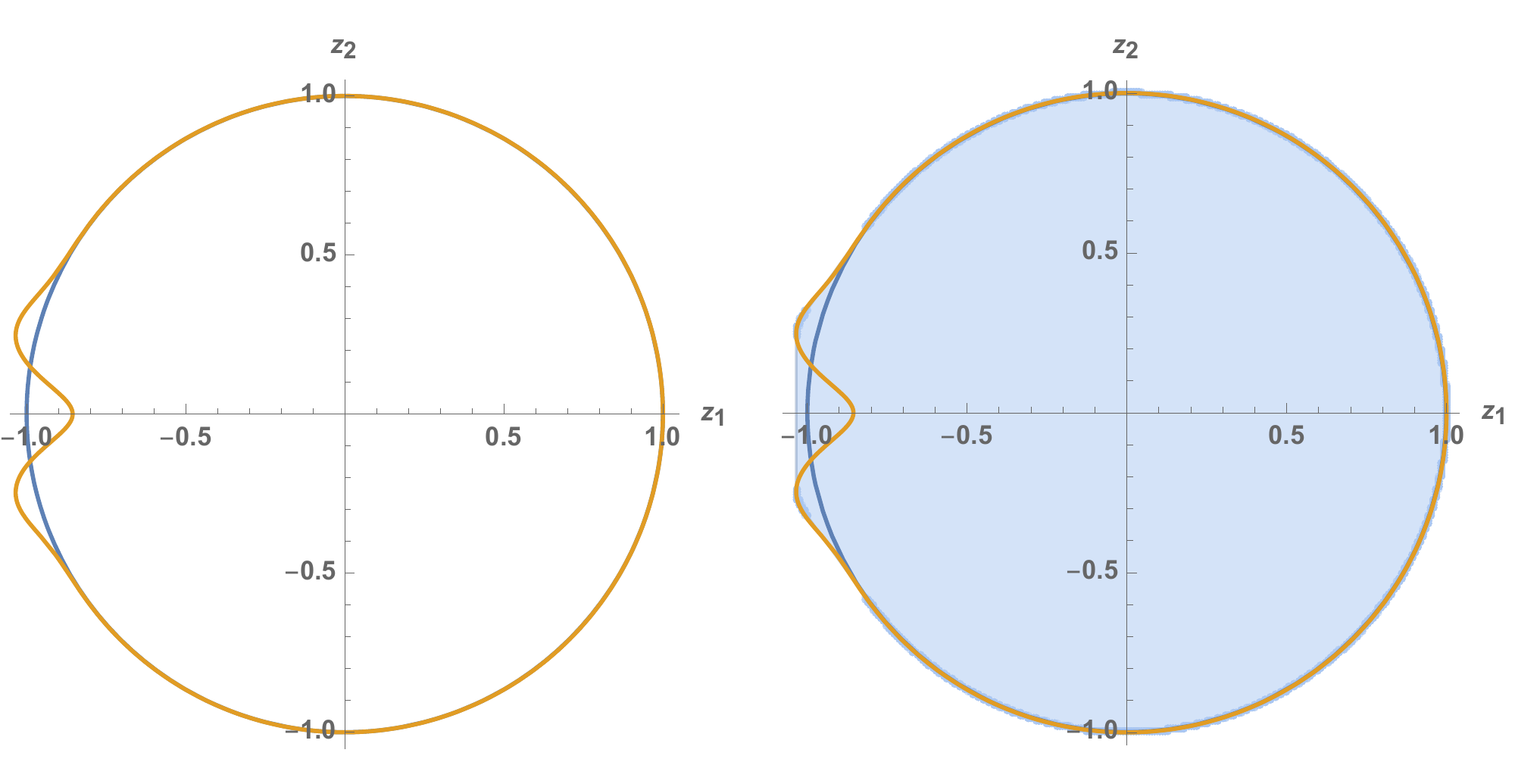}
\begin{flushleft}
\caption{(Left): an extremality curve that naively violates the WGC as it does not enclose the unit circle. (Right): the convex completion of the extremality curve \textit{does} enclose the unit circle, hence the WGC is satisfied. For this to be possible the extremality surface must be somewhere locally non-convex, which is shown in appendix \ref{appConvex} to be impossible in the perturbative regime.
\label{convex hull}}
\end{flushleft}
\end{figure}

This means that to determine if the decay of a large black hole is kinematically allowed, we must first determine the convex hull of a complicated surface, a task that may only be tractable numerically. As illustrated in figure \ref{convex hull}, it is possible for the convex hull of the extremality surface to enclose the unit ball even if the surface itself does not. Furthermore, the extremality surface may be non-convex even if the magnitude of the corrections is arbitrarily small. 

The condition simplifies somewhat in the $Q^2 \gg 1$ regime, where the corrections to the unit circle derive from the four-derivative terms and are small as a result. In appendix \ref{appConvex} we prove that if $T(\vec{z},Q^2)$ is a quartic form, as it is in the explicit result (\ref{extremality result}), then the smallness of the deviation \textit{does} imply convexity. In this regime, the convex hull condition is simplified in the sense that the extremality surface always bounds a convex region. At a given $Q^2 \gg 1$, and $\vec{z}$, the black hole extremality bound describes a surface in $z$-space of the form
\begin{equation}
    |\vec{z}|=1+\frac{1}{(Q^2)^3} T_{ijkl}z^i z^j z^k z^l + \mathcal{O}\left(\frac{1}{(Q^2)^2}\right).
\end{equation}
The condition for the multi-charge weak gravity conjecture to be satisfied in the perturbative regime degenerates to the more tractable condition:\\

\textbf{(Perturbative) Black Hole Weak Gravity Conjecture:} \textit{
It is kinematically possible for asymptotically large extremal black holes to decay into smaller finite $Q^2$ extremal black holes if the quartic extremality form} 
\begin{equation}
\label{extremalityform}
    T(q^i,p^i) = T_{ijkl}z^i z^j z^k z^l,
\end{equation}
 \textit{is everywhere non-negative. Using the parametrization of the effective action (\ref{Action}), this bound takes the form}
    \begin{align}
        T(q^i,p^i)\;\;= \;\;&2  \alpha_{ijkl}  (q^i q^j - p^i p^j)(q^k q^l - p^k p^l) +  8 \beta_{ijkl} q^i p^j q^k p^l - \gamma _{ij} \, Q^2\left( q^i q^j - p^i p^j\right)\nonumber \\
        &  \hspace{5mm} + 4 \,  \chi_{ijkl} \, q^i p^j \, \left(q^k q^l-p^k p^l\right) - 2 \,  \omega _{ij} \,  Q^2 q^i p^j \geq 0 \, ,
        \label{extremality bound}
\end{align}
\textit{which follows directly from (\ref{extremality result}).}

\subsection{Examples}

According to the previous section, we can determine whether black holes are stable by checking if the extremality form is anywhere negative. In this section we demonstrate this with a few basic examples. 

\subsubsection{Black Hole With Two Electric Charges}

A black hole that is electrically charged under two $U(1)$ groups provides one simple example. In this case, the extremality bound simplifies to 
\begin{align}
    (2 \alpha_{ijkl} - \gamma_{ij} \delta_{kl}) q^i q^j q^k q^l  > 0.
\end{align}
As the $q$ factors project to the completely symmetric part of this tensor, it is convenient to define $ T_{ijkl} = 2 \alpha_{ \{ijkl\} } - \gamma_{ \{ij} \delta_{kl \} }$, where we have symmetrized the indices with weight one. Expanding the constraint in components leads to 
\begin{align}
    \begin{split}
        T_{1111} \, q_1^4 + T_{1112} \, q_1^3 \, q_2 + T_{1122} \, q_1^2 \, q_2^2 + T_{1222} \, q_1 \, q_2^3  + T_{2222} \, q_2^4 > 0.
        \label{two electric polynomial}
    \end{split}
\end{align}
This polynomial must be positive for all possible combinations of $q_1$ and $q_2$. We use the fact that the polynomial in (\ref{two electric polynomial}) is homogenous, and divide by $(q_2)^4$. Redefining $q_1 / q_2 = x$ simplifies the left-hand-side of the inequality to a polynomial of one variable:
\begin{align}
    \begin{split}
        T_{1111} \, x^4 +  T_{1112} \, x^3 +  T_{1122} \, x^2 + T_{1222} \, x  + T_{2222} > 0.
    \end{split}
    \label{electric one var polynomial}
\end{align}
This polynomial is quartic so one may solve this by studying the explicit expressions for the roots and demanding that they are not real. However the positivity conditions for fourth order polynomials are much simpler and lead to a set of relations among the components of $T_{ijkl}$ (see, for instance, \cite{Wang2005CommentsO}). This allows the problem to be solved entirely in the case of two charges; for $N>2$ one must analyze multivariate polynomials. 

For an example of a theory that may be in the Swampland, consider the following four-derivative terms:
\begin{align}
    \begin{split}
        \mathcal{L}_4 = \alpha_{1111} \, F^1_{\mu \nu} F^{1 \, \mu \nu} F^1_{\rho \sigma} F^{1 \, \rho \sigma} + \alpha_{1122} \, F^1_{\mu \nu} F^{1 \, \mu \nu} F^2_{\rho \sigma} F^{2 \, \rho \sigma} + \alpha_{2222} \, F^2_{\mu \nu} F^{2 \, \mu \nu} F^2_{\rho \sigma} F^{2 \, \rho \sigma},
    \end{split}
\end{align}
where $\alpha_{1111} = 2$, $\alpha_{1122} = -8$, and $\alpha_{2222} = 3$. Then the extremality shift becomes
\begin{align}
    2 \, q_1^4 - 8 \, q_1^2 \, q_2^2 + 3 \, q_2^4 > 0.
\end{align}
The inequality is satisfied when $q_1 = 0$ or $q_2 = 0$, but at $q_1 = q_2$, the extremality shift is negative. Therefore, a black hole with $q_1 = q_2$ in this theory would not be able to decay to smaller black holes. This model requires the existence of self-repulsive states in the spectrum in either the particle or stringy regimes to evade the Swampland.

\subsubsection{Dyonic Black Hole}

Another simple case occurs when there is only a single gauge field but the black hole has both electric and magnetic charge. Then the extremality bound is obtained by removing all indices from (\ref{extremality result}):
\begin{align} 
    2 \alpha \, (q^2 - p^2)^2 + 8 \beta \, q^2 p^2 - \gamma \, (q^2 - p^2) (q^2 + p^2) + 4 \chi \, q p (q^2 - p^2)  - 2 \omega \, q p (q^2 + p^2) \ > \  0.
    \label{onedyonicext}
\end{align}
We recover the results of \cite{Kats:2006xp} when the magnetic charge is set to zero. A single electric charge shifts the extremality as 
\begin{equation}
    \label{singleelectric}
    |z_q| = 1 + \frac{2}{5 |Q|^2} ( 2 \alpha - \gamma ).
\end{equation}
However, a single magnetic charge has the opposite sign for $\gamma$:
\begin{equation}
    \label{singlemagnetic}
    |z_{p}| = 1 + \frac{2}{5 |Q|^2} ( 2 \alpha + \gamma ).
\end{equation}
Requiring that both types of black holes be able to decay places a stronger constraint on $\alpha$ and $\gamma$:
\begin{align}
    \begin{split}
        \label{combined}
        2 \alpha > |\gamma|.
    \end{split}
\end{align}
If we assume that both $p$ and $q$ are non-zero, we can again divide by $p^4$ as we did in the previous section, and again find a polynomial of a single variable:
\begin{align}
     (2 \alpha - \gamma) \, y^4 + (4 \chi - 2 \omega ) \, y^3 + (-4 \alpha  + 8 \beta) \, y^2 + (- 4 \chi - 2 \omega) \, y + (2 \alpha + \gamma) \ > \ 0.
     \label{dyon polynomial}
\end{align}

The generalized bound (\ref{dyon polynomial}) coincides exactly with the (regularized forward-limit) scattering positivity bounds derived in \cite{Bellazzini:2019xts} for arbitrary linear combinations of external states. It is interesting that the requirement that dyonic black holes are unstable gives a new physical motivation for these generalized scattering bounds.

For the case of a single gauge field, a very physical example comes to mind: the Euler-Heisenberg Lagrangian \cite{Heisenberg1936}, in which integrating out electron loops induces a four-point interaction among the gauge fields.\footnote{The electron should also contribute to the $W F F$-type operators as well, but this contribution is suppressed by a factor of $1/z$. The electron is extraordinarily superextremal ($z = 2 \times 10^{21}$) so we can safely ignore these terms for our example.} This model has four derivative terms given by
\begin{align}
    \mathcal{L}_4 = \alpha (F_{\mu \nu} F^{\mu \nu})^2 + \beta (F_{\mu \nu} \tilde{F}^{\mu \nu})^2,
\end{align}
with $\alpha = 4$, $\beta = 7$ (up to overall constants that do not effect the problem). The inequality that must be satisfied is the following:
\begin{align}
    4 y^4 + 40 y^2 + 8 > 0.
\end{align}
Clearly this holds for all values of $y$. Thus, we have found that the Euler-Heisenberg theory is not in the Swampland. This does not require that we know anything about the spectrum, or that the higher-derivative operators came from integrating out a particle at all. Only the four-derivative couplings are needed to learn that this theory allows nearly extremal black holes to decay.

The condition (\ref{onedyonicext}) exhibits an interesting simplification when $\alpha = \beta$ and the remaining coefficients are set to zero. In this case, the condition on the quartic form then reads
\begin{equation}
    \alpha (q^2+p^2)^2 > 0. 
\end{equation}
In this special case the extremality surface becomes invariant under orthogonal rotations in charge-space. In fact, it is simple to verify that this is the only choice of coefficients with this feature. The enhanced symmetry is a consequence of the electromagnetic duality invariance of the equations of motion for this choice of coefficients. In the effective action, the necessary condition for duality invariance is the \textit{Noether-Gaillard-Zumino condition} \cite{Gaillard:1981rj}
\begin{equation}
    F_{\mu\nu}\tilde{F}^{\mu\nu}+G_{\mu\nu}\tilde{G}^{\mu\nu} = 0, \hspace{5mm} \text{where} \hspace{5mm} \tilde{G}_{\mu\nu} \equiv 2 \frac{\delta S}{\delta F^{\mu\nu}}.
\end{equation}
One can verify that this is satisfied if we $\alpha = \beta, \ \gamma = \chi = \omega = 0$ as above, at least to fourth order in derivatives. To make this equation hold to sixth order would require the addition of sixth-derivative operators to the Lagrangian, and so on. For a general analysis of electric-magnetic duality invariant theories, see \cite{Gibbons:1995cv}. In the following section we show that the generalization of the electromagnetic duality group from $U(1)$ in the single charge case, to $U(N)$ in the $N$-charge case plays an essential role in renormalization group running of the four-derivative Wilson coefficients. 

\begin{figure}
\includegraphics[scale=0.8]{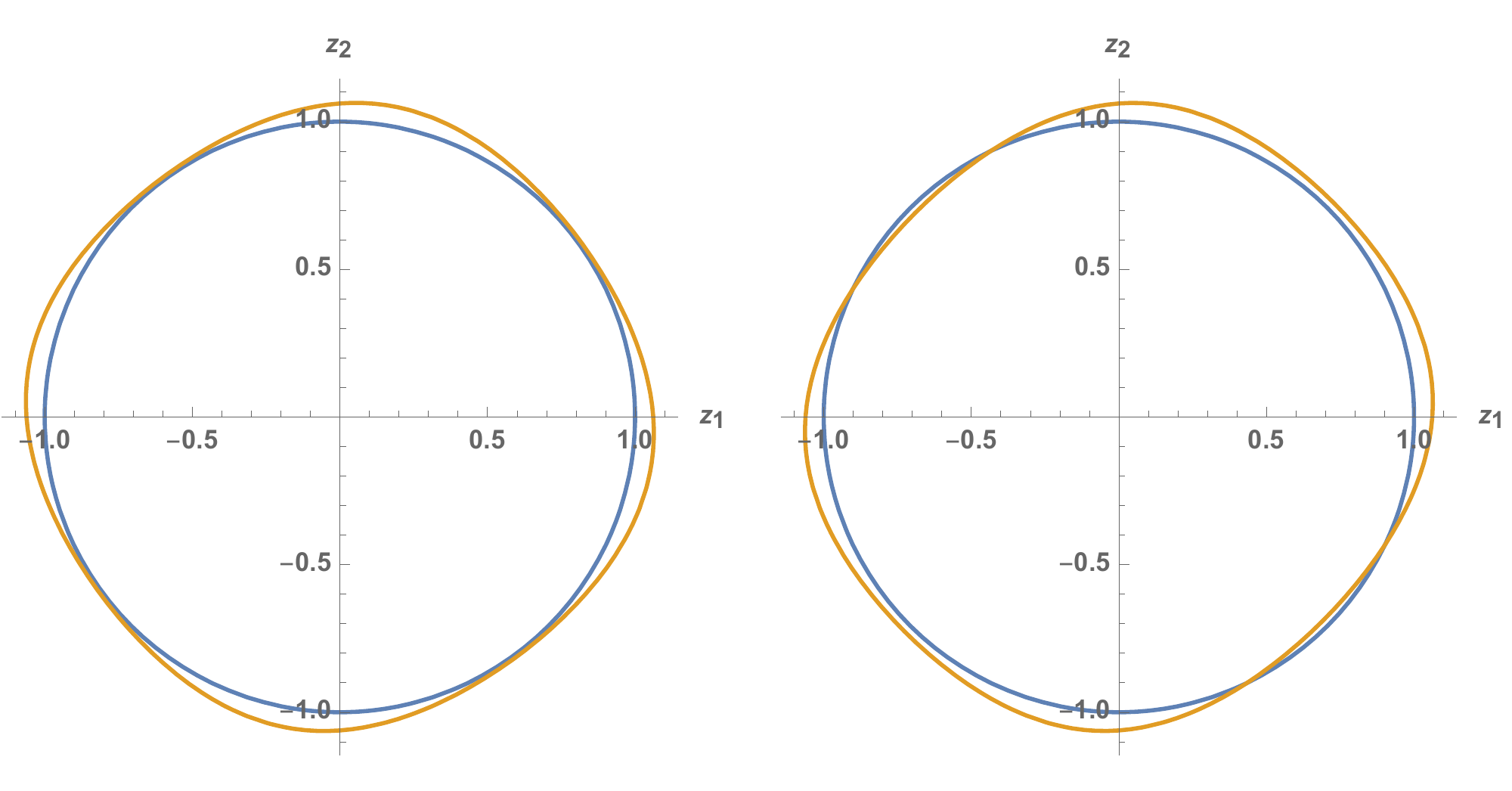}
\begin{flushleft}
\caption{
(Left): the corrections to the extremality curve are everywhere positive, hence the WGC is satisfied. (Right): the corrections to the extremality curve are \textit{not} everywhere positive; large extremal black holes cannot always decay to intermediate mass black holes, whether or not the WGC is satisfied cannot be decided in the low-energy EFT.
\label{positive corrections}}
\end{flushleft}
\end{figure}

\subsection{Unitarity and Causality}

Infrared consistency conditions on the low energy effective theory have been used to bound the coefficients of higher-derivative operators. Such constraints were first considered in the context of the weak gravity conjecture in \cite{Cheung:2014ega}, and were extended to the case of multiple gauge fields in \cite{Andriolo:2018lvp}. Further arguments based on unitarity and causality were given in \cite{Hamada:2018dde}. Here we review these arguments and present a few generalizations.

\subsubsection{Integrating Out Massive Particles}

One source of higher derivative corrections derives from integrating out states in the particle regime. By this we mean states that are well described by ordinary QFT on a fixed spacetime background. Such states necessarily have masses smaller than some cutoff scale $\Lambda_{QFT}$, which is the string scale or whatever scale new physics invalidates the QFT description. We have already seen a simple example of this in the Euler-Heisenberg Lagrangian above.

At tree-level, only neutral particles contribute to the four-point interactions. Consider, for example, a dilaton that couples to the field strengths. The Lagrangian for the scalar theory is
\begin{align}
    \mathcal{L} = \frac{R}{4} - \frac{1}{2} ( \partial \phi )^2 - \frac{m_{\phi}^2}{2} \phi^2 - \frac{1}{4} F^i_{\mu \nu} F^{i \, \mu \nu} + \mu_{i j} \phi F^i_{\mu \nu} F^{j \, \mu \nu}.
\end{align}
We integrate out the scalar to find the effective four-derivative coupling by matching to the low-energy EFT at the scale $\Lambda_{\text{UV}}\lesssim m_\phi$
\begin{align}
    \mathcal{L}_4 \supset \frac{M_{\text{Pl}}^4}{m_{\phi}^2}  ( \mu_{ij} \mu_{kl} + \mu_{ik} \mu_{jl} + \mu_{il} \mu_{jk} ) F^i_{\mu \nu} F^{j \, \mu \nu} F^k_{\rho \sigma} F^{l \, \rho \sigma}.
\end{align}
Therefore, in this simple setup, the coefficient $\alpha_{ijkl}$ takes the form
\begin{align}
    \alpha_{ijkl} =  \frac{1}{m_{\phi}^2}  ( \mu_{ij} \mu_{kl} + \mu_{ik} \mu_{jl} + \mu_{il} \mu_{jk} ).
\end{align}

For a single gauge field $\alpha =  \frac{3 \mu^2}{m_{\phi}^2} $. Unitarity requires that $\mu$ is real, which implies that $\alpha$ is positive \cite{Hamada:2018dde}. It is easy to see that this is still the case when there are more gauge fields. The extremality form for this theory is 
\begin{align}
    \alpha_{ijkl} q^i q^j q^k q^l = \frac{3}{m_{\phi}^2} (\mu_{ij} q^i q^j)^2,
\end{align}
which must be positive.\footnote{Note that unlike the case of single gauge field, unitarity does not bound all the coefficients separately. For instance, in the two charge case, $\mu_{11} = 1$, $\mu_{22} = -1$, and $\mu_{12} = 0$ would lead to $\alpha_{1122} = -1/m_{\phi}^2$.} The same reasoning shows that integrating out an axion, which couples to $F^i \tilde{F}^j$, generates $\beta_{ijlk}$, and that its contribution to the extremality form is also positive.

Light charged particles cannot contribute at tree-level so their leading contributions are at loop-level. The diagrams that contribute in this case are:
\begin{center}
	\begin{tikzpicture}[scale=0.9, line width=1 pt]
	\begin{scope}
\draw[vector] (0,0)--(-1,1);
\draw[scalar] (0,0)--(2,0);
\draw[scalarbar] (0,0)--(0,-2);
\draw[scalar] (2,0)--(2,-2);
\draw[scalarbar] (0,-2)--(2,-2);
\draw[vector] (2,0)--(3,1);
\draw[vector] (0,-2)--(-1,-3);
\draw[vector] (2,-2)--(3,-3);
	\node [above left] at (-1,1) {$\gamma_i$};
	\node [below left] at (-1,-3) {$\gamma_j$};
	\node [above right] at (3,1) {$\gamma_l$};
	\node [below right] at (3,-3) {$\gamma_k$};
	\node at (1,-4) {(a)};
	\end{scope}
	\begin{scope}[shift={(5,-1)}]
	\draw [vector] (2,2)--(3,1);
	\draw [scalar] (3,1)--(4,0);    
	\draw [vector] (2,-2)--(3,-1);	
	\draw [scalarbar] (3,1)--(3,-1);
	\draw [scalarbar] (3,-1)--(4,0);    
	\draw [vector] (4,0.05)--(5,0.05);	
	\draw [vector] (4,-0.05)--(5,-0.05);	
	\draw [vector] (5,0)--(6,1);
	\draw [vector] (5,0)--(6,-1);
	\node [above left] at (2,2) {$\gamma_i$};
	\node [below left] at (2,-2) {$\gamma_j$};
	\node [above right] at (6,1) {$\gamma_l$};
	\node [below right] at (6,-1) {$\gamma_k$};
	\node at (4,-3) {(b)};
	\end{scope} 
	\end{tikzpicture}
\end{center}
\begin{center}
    \begin{tikzpicture}[scale=0.9, line width=1 pt]
    \begin{scope}[shift={(0,0)}]
	\draw [vector] (-1,1)--(0,0);
	\draw [vector] (-1,-1)--(0,0);
	\draw [vector] (0,0.05)--(1.1,0.05);
	\draw [vector] (0,-0.05)--(1.1,-0.05);	
	\draw [scalar] (3,0) arc (0:180:1);
	\draw [scalarbar] (3,0) arc (0:-180:1); 
	\draw [vector] (3,0.05)--(4,0.05);	
	\draw [vector] (3,-0.05)--(4,-0.05);	
	\draw [vector] (4,0)--(5,1);
	\draw [vector] (4,0)--(5,-1);
	\node [above left] at (-1,1) {$\gamma_i$};
	\node [below left] at (-1,-1) {$\gamma_j$};
	\node [above right] at (5,1) {$\gamma_l$};
	\node [below right] at (5,-1) {$\gamma_k$};
	\node at (2,-3) {(c)};
	\end{scope} 
	\begin{scope}[shift={(7,0)}]
	\draw [vector] (2,2)--(3,1);
	\draw [scalar] (3,1)--(4.5,0);    
	\draw [vector] (2,-2)--(3,-1);	
	\draw [scalarbar] (3,1)--(3,-1);
	\draw [scalarbar] (3,-1)--(4.5,0);    
	\draw [vector] (4.5,0.05)--(6,0.05);	
	\draw [vector] (4.5,-0.05)--(6,-0.05);	
	\node [above left] at (2,2) {$\gamma_i$};
	\node [below left] at (2,-2) {$\gamma_j$};
	\node [right] at (6,0) {$h$};
	\node at (4,-3) {(d)};
	\end{scope}
    \end{tikzpicture}
\end{center}
These contribute at the same order except they have relative factors of $z_{\phi}$, the particle's charge-to-mass ratio, coming from counting couplings and propagators. Diagram (a) goes like $z_{\phi}^4$, (b) like $z_{\phi}^2$, (c) like $z_{\phi}^0$; diagram (d) contributes at order $z_{\phi}^2$.  The field-strength four-point interaction is generated by the first three diagrams. In the limit where $z_{\phi} \gg 1$, diagram (a) dominates all the others (as we noted above in the Euler-Heisenberg example) and the extremality form becomes
\begin{align}
    T_{ijlk} q^i q^j q^k q^l= \alpha_{ijkl} q^i q^j q^k q^l = (z_{\phi}^i q^i)^4,
\end{align}
Again, we find a manifestly positive contribution. For $z_{\phi} $ near or less than one, both $\alpha_{ijkl}$ and $\gamma_{ij}$ are generated by diagrams that are order $z_{\phi}^0$. In that case this scaling argument does not apply, and the order one constants need to be included in the analysis. These arguments are schematic and largely review what was already considered in \cite{Andriolo:2018lvp}.

One might wonder whether this analysis is relevant to the parity-odd operators. Interestingly, \cite{Colwell:2015wna} has shown how to generalize the Euler-Heisenberg Lagrangian by integrating out a monopole or dyonic charge. The effective Lagrangian was derived in that paper (and earlier in \cite{Kovalevich:1997de}) to be
 \begin{align}
     \begin{split}
     \mathcal{L}_4 =&  \big( 4 (\hat{q}^2 - \hat{p}^2)^2 + 28\hat{q}^2 \hat{p}^2 \big) (F^2)^2
    + \big( 7 (\hat{q}^2 - \hat{p}^2)^2 +  16 \hat{q}^2 \hat{p}^2 \big) (F \tilde{F} )^2 - 12 \hat{q} \hat{p}  (\hat{q}^2 - \hat{p}^2) F^2 (F \tilde{F}).
    \label{monopole}
     \end{split}
 \end{align}
where the $\hat{q}$ and $\hat{p}$ refer to the electric and magnetic charges of the dyon that is integrated out (not the charges of the black hole). This procedure generates the parity-violating four-photon coupling as well as the two parity-even ones. This is not surprising given that magnetic charges violate parity in their interactions with the gauge field. What is more interesting is that this term is \textit{not} a square, unlike every other term appearing in the effective Lagrangian. The sign of the generated term depends on the sign of the product of the electric and magnetic charges of the particle. In terms of the polynomial derived in (\ref{dyon polynomial}), the condition that must be met to satisfy the WGC is:
\begin{align}
    \begin{split}
        &  \big( \hat{q}^4 + 5 \hat{q}^2 \hat{p}^2 + \hat{p}^4 \big)\, x^4
        + \, 3 \, \big( \hat{q}^3  \hat{p} - \hat{q}  \hat{p}^3 \big) \, x^3
        + \,  \big( 5 \hat{q}^4 -8 \hat{q}^2 \hat{p}^2 + 5 \hat{p}^4  \big) \, x^2 \\
        & \qquad \qquad  + \, 3  \, \big( \hat{q}^3  \hat{p} - \hat{q}  \hat{p}^3 \big) \, x 
        \,+ \,  \big( \hat{q}^4 + 5 \hat{q}^2 \hat{p}^2 + \hat{p}^4 \big)  \, > \, 0.
    \end{split}
\end{align}
This polynomial is always positive, so the Lagrangian given in (\ref{monopole}) does not allow for stable black holes and satisfies the WGC.
 
\subsubsection{Causality Constraints}

Another set of arguments for bounds on the EFT coefficients rely on causality. These were first considered in \cite{Cheung:2014ega} and generalized to multiple gauge fields in \cite{Hamada:2018dde}. Two methods were used, and they were shown to give the same result. The first is to consider the propagation of photons on a photon gas background. Requiring that photons travel do not travel superluminally constrains the four-photon interaction. The second method uses analyticity and unitarity to relate the EFT coefficients to an integral over the imaginary part of the amplitude, which is manifestly positive. The bounds obtained this way for multiple gauge fields are
\begin{align}
    \label{causality}
    \sum_{ij} \left( \alpha_{ \{i j \} \{k l \} } + \beta_{ \{i j \} \{k l \} }  \right)  u^i v^j u^k v^l \geq 0.
\end{align}
This inequality must hold for any vectors $\vec{u}$ and $\vec{v}$. This bound is independent from the bounds that we have derived in (\ref{extremality bound}), so it is not enough to imply the WGC on its own. 

So far these arguments have only bounded the four-photon interactions. Another causality-based argument was made in \cite{Hamada:2018dde} that bounds the photon-photon-graviton interaction parameterized by $\gamma$. They argued that the addition of this four-derivative term introduces causality violation at a scale $E \sim M_{\text{Pl}} / \gamma^{1/2}$ (a fact noticed in \cite{Camanho:2014apa}). Therefore new physics must arise at scale $\Lambda_{QFT} \lesssim M_{\text{Pl}} / \gamma^{1/2}$, which means $\gamma \lesssim (M_{\text{Pl}} / \Lambda_{QFT})^2$. This argument suggests that perhaps the $W F F$ four-derivative terms are generically bounded by causality to be much smaller than a number of possible contributions to the $F^4$ terms. It would be interesting to extend the analysis of \cite{Camanho:2014apa} to the more general set of operators used here, but this is beyond the scope of our paper. 

\section{Renormalization of Four-Derivative Operators}
\label{sec renorm}

The Wilson coefficients that appear in the extremality shift (\ref{extremality result}) are determined by UV degrees-of-freedom integrated out of the low-energy effective field theory. In section \ref{sec:decay} we gave explicit examples of contributions to the Wilson coefficients from integrating out massive particle states, both at tree- and loop-level. To consistently calculate the correction to the extremality bound for a black hole with total charge $Q^2$, we must first calculate the renormalization group evolution from the \textit{matching scale} $\mu^2 \sim \Lambda_{\text{UV}}^2$ to the \textit{horizon scale} $\mu^2 \sim M_{\text{Pl}}^2/Q^2$. For black holes with $Q^2\gg 1$ these scales can be arbitrarily separated and the effects of the logarithmic running of the Wilson coefficients can be dramatic. 

In the single $U(1)$ case it was recently argued \cite{Charles:2019qqt} that as we RG flow towards the deep IR, $Q^2\rightarrow \infty$, the logarithmic running of a particular combination of Wilson coefficients dominates the extremality shift, independent of the values of the coefficients at the matching scale. Explicitly, the extremality bound takes the form
\begin{equation}
\frac{Q^2}{M^2}\leq 1 + \frac{4}{5Q^2}\left(\frac{c}{16\pi^2}\text{log}\left(\frac{\Lambda_{\text{UV}}^2 Q^2}{M_{\text{Pl}}^2}\right)+2\alpha_{\text{UV}}-\gamma_{\text{UV}}\right).
\end{equation}
If $c>0$ then at some finite value of the charge $Q^2$ extremal black holes must be self-repulsive. This was shown to be the case in \cite{Charles:2019qqt} for various explicit theories, including the single $U(1)$ model (\ref{4dersingle}). Since the renormalization group coefficient $c$ depends only on the massless degrees of freedom, this analysis depends only on the universality class of the model. For those classes in which this conclusion holds, the WGC is always satisfied independently of the details of the UV completion, and in that sense is no longer a useful Swampland criterion.

In this section we show how this argument generalizes to an arbitrary number of $U(1)$ gauge fields. Since there are many more four-derivative operators, we emphasize the importance of a non-renormalization theorem that arises as a consequence of the accidental $U(N)$ electromagnetic duality symmetry of the two-derivative approximation. In the following subsection we give an on-shell proof of this theorem, and then use it to extend the argument above. 

\subsection{Non-Renormalization and Electromagnetic Duality}

Consider a low-energy effective action of the form
\begin{equation}
\label{emnaction}
    S = \int\text{d}^4x\sqrt{-g}\left[\frac{1}{4}R - \frac{1}{4} F_{\mu\nu}^i F^{i\mu\nu} + \sum_i c_i \mathcal{O}_i\right],
\end{equation}
where the operators $\mathcal{O}_i$ have at least three derivatives. The Wilson coefficient $c_i$ is \textit{renormalized} if $\mathcal{O}_i$ corresponds to a counterterm to an ultraviolet divergence. In terms of on-shell scattering amplitudes, an operator $\mathcal{O}_i$ generates an on-shell local matrix element, and the coefficient $c_i$ is renormalized if there is a corresponding one-loop scattering amplitude with an ultraviolet divergence. Here ``corresponding" means that the external states of the matrix element of $\mathcal{O}_i$ must agree with the external states of the loop amplitude. Conversely, an operator $\mathcal{O}_i$ is \textit{not} renormalized if there are no corresponding UV divergent one-loop scattering amplitudes with the correct external states. 

We begin by making the observation that the leading, two-derivative, part of the action (\ref{emnaction}) has an accidental $O(N)$ flavor symmetry. This leads to a rather trivial non-renormalization theorem:\\

\textit{In Einstein-Maxwell with $N$ $U(1)$ gauge fields, a four-derivative operator $\mathcal{O}_i$ is renormalized at one-loop only if it generates an on-shell local matrix element that is an invariant tensor of the flavor symmetry group $O(N)$.}\\

This statement is trivial because there are no Feynman diagrams at one-loop that are not $O(N)$-invariant. Since we are not assuming that $O(N)$ is a symmetry of the UV completion, such symmetry violating higher-derivative operators may appear in the effective action, but they cannot act as counterterms to ultraviolet divergences, and hence their associated Wilson coefficients do not have a logarithmic running. Trivial non-renormalization theorems of this kind follow for all symmetries of the effective action. 

The non-trivial non-renormalization theorem we prove below concerns \textit{electromagnetic duality symmetries}, which are only symmetries of the equations of motion, not the action \cite{Gaillard:1981rj}. Consequently, they are not manifest off-shell, meaning diagram-by-diagram in the standard covariant Feynman diagram expansion, and the above reasoning is no longer valid. Nonetheless we will prove that the above non-renormalization theorem is valid verbatim, at least at one-loop, where the flavor symmetry group $O(N)$ is enhanced to the maximal compact electromagnetic duality group $U(N)$. 

It is convenient to discuss UV divergences in the context of dimensional regularization where the loop integration is performed in $d=4-2\epsilon$ dimensions and ultraviolet divergences at one-loop appear as $1/\epsilon$ poles. In this context we can classify the sources of UV divergences in on-shell scattering amplitudes:

\begin{enumerate}
    \item \textbf{Cut-Constructible Divergences}: By standard integral reduction algorithms, one-loop amplitudes admit a universal decomposition into a sum over a set of \textit{master integrals}:
\begin{equation}
    \label{loopdecomp}
    \mathcal{A}_n^{\text{1-loop}} = \sum_{i}a_i I_i^{(\text{box})}+\sum_j b_j I_j^{(\text{triangle})} + \sum_k c_k I_k^{(\text{bubble})} + \mathcal{R},
\end{equation}
where the master integrals are \textit{scalar} integrals with the indicated topology. Here $a_i,b_j,c_k$ and $\mathcal{R}$ are rational functions of the external kinematic data. The first three contributions are often referred to as the \textit{cut-constructible} part of the amplitude; they contain all of the branch cut discontinuities required by perturbative unitarity at one-loop. These contributions can be completely determined from on-shell unitarity cuts into physical tree amplitudes \cite{Britto:2010xq,Bern:1994cg}. This determines the one-loop amplitude up to a rational ambiguity indicated by $\mathcal{R}$. Since the rational part is both UV and IR finite, the divergent structure (both UV and IR) of the one-loop amplitude is completely determined by the tree-level scattering amplitudes. From the definition it is clear that only the master bubble integral $I^{\text{bubble}}$ is UV divergent, 
\begin{equation}
    \left[I^{\text{bubble}}\left(K^2\right)\right]_{\text{UV}} \equiv \left[\int\frac{\text{d}^{4-2\epsilon}l}{(2\pi)^{4-2\epsilon}} \frac{1}{l^2(l-K)^2}\right]_{\text{UV}} = \frac{i}{16\pi^2 \epsilon},
\end{equation}
and therefore what we call the \textit{cut-constructible divergence} is proportional to the sum of the bubble coefficients $c_k$. These coefficients are completely determined by the two-particle unitarity cuts of the one-loop amplitude. It has been shown that the two-particle unitarity cuts of the master bubble integrals are purely rational functions, while the two-particle cuts of the triangle and box integrals give logarithms \cite{Britto:2010xq,Bern:1994cg}. By explicitly calculating the two-particle cuts of $\mathcal{A}_n^{\text{1-loop}}$ one can read off the rational part as the associated bubble coefficient. Using the relation between unitarity cuts of one-loop amplitudes and on-shell phase space integrals of tree-amplitudes gives a well-known general formula for the cut-constructible UV divergence
\begin{equation}
\label{cutdiv}
    [\mathcal{A}_n^{\text{1-loop}}]_{\text{UV}}=\sum_{\text{cuts}}\left[\int \text{d}\mu_{\text{LIPS}}\sum_{\text{states}}\mathcal{A}^{\text{tree}}_L\mathcal{A}^{\text{tree}}_R\right]_{\text{Rational}},
\end{equation}
where the sums on the right-hand-side are taken over all cuts \textit{and} all on-shell states exchanged in each cut. The details of the integration in this formula are not essential to the argument we make below.
\item \textbf{UV/IR Mixed Divergences}: In dimensional regularization IR divergences are also regularized as $1/\epsilon$ poles. Even though their physical origin is very different there can be non-trivial cancellations between UV and IR divergences in the on-shell scattering amplitude. Such \textit{mixed} UV divergences are just as important as the cut-constructible ones, and must be included to calculate the correct beta functions \cite{ArkaniHamed:2008gz,Huang:2012aq}. Unfortunately, due to this cancellation they cannot be immediately extracted from the cut-constructible part of the one-loop amplitude (\ref{loopdecomp}). The strategy is to first independently determine the expected one-loop IR divergence, and then compare against the IR divergences in the cut-constructible part of the amplitude. Any discrepancy must be due to UV/IR cancellations, and so can be used to infer the mixed UV divergences. The true IR divergent structure is determined by the KLN theorem \cite{Kinoshita:1962ur}. This states that in an inclusive cross-section, virtual IR divergences from loop integration must cancel against divergences in the initial/final phase space integrals that arise from soft/collinear real emission. Such real emission singularities are fixed by tree-level soft/collinear limits, so we find that again the mixed divergences are completely reconstructible from tree-level, physical data. 
\end{enumerate}

We begin with an on-shell description of $U(N)$ duality invariance at tree-level. The three-particle amplitudes are completely fixed \footnote{The spinor-helicity conventions used in these expressions are given in \cite{Elvang:2015rqa}.}:
\begin{align}
    &\mathcal{A}^{\text{tree}}_3\left(1_h^+,2_h^+,3_h^-\right) = \frac{[12]^6}{[23]^2[31]^2}, \hspace{5mm} \mathcal{A}^{\text{tree}}_3\left(1_h^-,2_h^-,3_h^+\right) = \frac{\langle 12\rangle^6}{\langle 23\rangle^2\langle 31\rangle^2}, \nonumber\\
    &\mathcal{A}^{\text{tree}}_3\left(1_h^+,2_{\gamma,i}^+,3_{\gamma}^{-,j}\right) = {\delta_i}^j\frac{[12]^4}{[23]^2}, \hspace{5mm} \mathcal{A}^{\text{tree}}_3\left(1_h^-,2_{\gamma,i}^+,3_{\gamma}^{-,j}\right) = {\delta_i}^j\frac{\langle 13\rangle^4}{\langle 23\rangle^2},
\end{align}
where $i,j=1,...,N$ are flavor indices. The fact that the on-shell three-particle amplitudes are diagonal in flavor space with unit coupling to the graviton is an on-shell expression of the Einstein equivalence principle. $U(N)$ duality invariance is encoded in the on-shell Ward identity:
\begin{equation}
    {U_i}^k{U^{*j}}_l\mathcal{A}^{\text{tree}}_3\left(1_h^+,2_{\gamma,k}^+,3_{\gamma}^{-,l}\right) = \mathcal{A}^{\text{tree}}_3\left(1_h^+,2_{\gamma,i}^+,3_{\gamma}^{-,j}\right),
\end{equation}
where $U\in U(N)$. In the explicit expressions above this is seen to hold as a consequence of the fact that ${\delta_i}^j$ is a $U(N)$-invariant tensor. The 4-point amplitudes are simple to calculate using on-shell recursion  
\begin{align}
    &\mathcal{A}_4^{\text{tree}}\left(1_h^+,2_h^+,3_h^-,4_h^-\right) = \frac{[12]^4\langle 34\rangle^4}{s_{12}s_{13}s_{14}}, \hspace{5mm} \mathcal{A}_4^{\text{tree}}\left(1_h^+,2_{\gamma, i}^+,3_h^-,4_{\gamma}^{-,j}\right) = {\delta_i}^j\frac{[12]^4\langle 34\rangle^2\langle 23\rangle^2}{s_{12}s_{13}s_{14}},\nonumber\\
    &\hspace{20mm} \mathcal{A}_4^{\text{tree}}\left(1_{\gamma,i}^+,2_{\gamma, j}^+,3_\gamma^{-,k},4_{\gamma}^{-,l}\right) = {\delta_i}^k{\delta_j}^l\frac{[12]^2\langle 34 \rangle^2}{s_{13}}+{\delta_i}^l{\delta_j}^k\frac{[12]^2\langle 34 \rangle^2}{s_{14}}.
\end{align}
Again, each of these is a $U(N)$-invariant tensor. As we discussed above, in the standard Lorentz covariant Feynman diagrammatic approach, only the $O(N)$ subgroup of global flavor rotations is manifest. The enhancement to the full $U(N)$ duality invariance in the on-shell amplitudes appears miraculous. A simple way to see that this enhancement continues to all multiplicities is to calculate the tree-amplitudes using on-shell recursion. Here the amplitude is given as a sum over factorization channels of the form
\begin{equation}
    \label{recursion}
    \mathcal{A}_n^{\text{tree}} \sim \sum_{\text{channels}} \sum_{\text{states}} \mathcal{A}^{\text{tree}}_L \mathcal{A}^{\text{tree}}_R.
\end{equation}
The precise details of the formula are not important to the argument. It is straightforward to prove $U(N)$ invariance by induction. Assume that all tree amplitudes $\mathcal{A}^{\text{tree}}_m$, with $m<n$ are duality invariant; using the recursive representation (\ref{recursion}) we show that $\mathcal{A}^{\text{tree}}_n$ is duality invariant channel-by-channel. If the exchanged on-shell state in a given channel is a graviton, then $\mathcal{A}_L^{\text{tree}}\mathcal{A}_R^{\text{tree}}$ is a product of invariant tensors, and hence invariant. If the exchanged state is a photon then the sum over helicity and the flavor index takes the form
\begin{align}  
    \label{photonexchange}
    &\mathcal{A}^{\text{tree}}_L\left(...,-p_{\gamma,i}^+\right)\mathcal{A}^{\text{tree}}_R\left(p_{\gamma}^{-,i},...\right) + \mathcal{A}^{\text{tree}}_L\left(...,-p_{\gamma}^{-,i}\right)\mathcal{A}^{\text{tree}}_R\left(p_{\gamma,i}^+,...\right).
\end{align}
Since this is the contraction of two tensors by the invariant ${\delta_i}^j$, it follows that this sum is likewise an invariant. Together with the explicitly verified duality invariance of the three-point amplitudes, the all-multiplicity Ward identity follows by induction. Here the key property we used was the existence of a valid on-shell recursion for the tree-level S-matrix (\ref{recursion}); a general discussion the necessary conditions for this to exist can be found in \cite{Cheung:2015ota}. 

We are now ready to prove the following non-renormalization theorem: \\

\textbf{Non-Renormalization of Duality Violating Operators}: \textit{In Einstein-Maxwell with $N$ $U(1)$ gauge fields, a four-derivative operator $\mathcal{O}_i$ is renormalized at one-loop only if it generates an on-shell local matrix element that is an invariant tensor of the maximal compact electromagnetic duality group $U(N)$.}\\

This result was first noted long-ago following a detailed calculation of the UV divergence \cite{Deser:1974cz,Deser:1974xq}, and recently generalized (including massless scalars) to the full non-compact duality group $Sp(2N)$ in \cite{Charles:2017dbr}. The new result in this section is a simple argument that demonstrates the duality invariance of the divergence \textit{without} the need for a detailed calculation.

We will prove that the total UV divergence is given by a sum over $U(N)$ invariant tensors. Beginning with the cut-constructible part, the logic here is very similar to the inductive proof of tree-level invariance. We will show that the divergence is a $U(N)$ invariant tensor cut-by-cut. In the representation (\ref{cutdiv}) we consider the contribution of a single two-particle cut; this can be either graviton-graviton, graviton-photon or photon-photon:

\begin{center}
	\begin{tikzpicture}[scale=0.9, line width=1 pt]
	\begin{scope}
	\draw [scalarnoarrow] (-1.5,1)--(0,0);
	\node at (-1.5,0) {$\huge\vdots$};
	\draw [scalarnoarrow] (-1.5,-1)--(0,0);
	\draw [vector] (0.1,0) arc (180:0:0.9);
	\draw [vector,segment length=3.9mm] (0,0) arc (180:0:1);
	\draw [vector] (0.1,0) arc (180:360:0.9);
	\draw [vector,segment length=3.9mm] (0,0) arc (180:360:1);   
	\draw [scalarnoarrow] (2,0)--(3.5,1);
	\draw [scalarnoarrow] (2,0)--(3.5,-1);
	\node at (3.5,0) {$\huge\vdots$};
	\draw[black,fill=white] (0,0) circle (3ex);
	\draw[black,fill=white] (2,0) circle (3ex);
	\node at (0,0) {\small $\mathcal{A}_L^\text{tree}$};
	\node at (2,0) {\small $\mathcal{A}_R^\text{tree}$};
	\draw[red,dashed] (1,-1.5)--(1,1.5);
	\end{scope}
	\begin{scope}[shift={(6,0)}]
	\draw [scalarnoarrow] (-1.5,1)--(0,0);
	\node at (-1.5,0) {$\huge\vdots$};
	\draw [scalarnoarrow] (-1.5,-1)--(0,0);
	\draw [vector] (0.1,0) arc (180:0:0.9);
	\draw [vector] (0.1,0) arc (180:360:0.9);
	\draw [vector,segment length=3.9mm] (0,0) arc (180:360:1);   
	\draw [scalarnoarrow] (2,0)--(3.5,1);
	\draw [scalarnoarrow] (2,0)--(3.5,-1);
	\node at (3.5,0) {$\huge\vdots$};
	\draw[black,fill=white] (0,0) circle (3ex);
	\draw[black,fill=white] (2,0) circle (3ex);
	\node at (0,0) {\small $\mathcal{A}_L^\text{tree}$};
	\node at (2,0) {\small $\mathcal{A}_R^\text{tree}$};
	\draw[red,dashed] (1,-1.5)--(1,1.5);
	\node at (0.2,1.2) {$\gamma^+_i$};
	\node at (1.9,1.2) {$\gamma^{-i}$};  
	\end{scope} 
	\begin{scope}[shift={(12,0)}]
	\draw [scalarnoarrow] (-1.5,1)--(0,0);
	\node at (-1.5,0) {$\huge\vdots$};
	\draw [scalarnoarrow] (-1.5,-1)--(0,0);
	\draw [vector] (0.1,0) arc (180:0:0.9);
	\draw [vector] (0.1,0) arc (180:360:0.9);
	\draw [scalarnoarrow] (2,0)--(3.5,1);
	\draw [scalarnoarrow] (2,0)--(3.5,-1);
	\node at (3.5,0) {$\huge\vdots$};
	\draw[black,fill=white] (0,0) circle (3ex);
	\draw[black,fill=white] (2,0) circle (3ex);
	\node at (0,0) {\small $\mathcal{A}_L^\text{tree}$};
	\node at (2,0) {\small $\mathcal{A}_R^\text{tree}$};
	\draw[red,dashed] (1,-1.5)--(1,1.5);
	\node at (0.2,1.2) {$\gamma^+_i$};
	\node at (1.9,1.2) {$\gamma^{-i}$}; 
	\node at (0.2,-1.2) {$\gamma^+_j$};
	\node at (1.9,-1.2) {$\gamma^{-j}$}; 
	\end{scope} 
	\end{tikzpicture}
\end{center}

Since the tree-amplitudes are invariant, and as in the expression (\ref{photonexchange}) the exchanged photon flavor indices are contracted with invariant tensors, each case separately generates an invariant tensor. Summing over all states and cuts we conclude that the cut-constructible divergence is duality invariant.

As for the possible mixed divergence, here we begin with the full IR divergence at one-loop. This is given by the universal formula
\cite{Dunbar:1995ed}
\begin{equation}
    \label{fullIR}
    \left[\mathcal{A}_n^{\text{1-loop}}\right]_{\text{IR}} =  \frac{ir_\Gamma }{(4\pi)^{2-\epsilon}} \frac{1}{\epsilon^2}\mathcal{A}_n^{\text{tree}}\sum_{i\neq j}^n (s_{ij})^{1-\epsilon},
\end{equation}
where the tree-amplitude on the right-hand-side and the loop amplitude on the left-hand-side have the same external states and $r_\Gamma = \Gamma^2(1-\epsilon)\Gamma(1+\epsilon)/\Gamma(2-\epsilon)$. As discussed above, in general there may be non-trivial UV/IR cancellations in the cut-constructible part of the one-loop amplitude. These can be disentangled using knowledge of the full IR divergence. In this case, things are somewhat simpler, and expanding the final factor in (\ref{fullIR}) gives
\begin{equation}
    \sum_{i\neq j}^n (s_{ij})^{1-\epsilon} = \sum_{i\neq j}^n s_{ij}+\epsilon\sum_{i\neq j}^n s_{ij}\log\left(s_{ij}\right) +\mathcal{O}\left(\epsilon^2\right).
\end{equation}
The first term in this sum is zero by momentum conservation. Expanding (\ref{fullIR}) the full IR divergence has the form 
\begin{equation}
    \left[\mathcal{A}_n^{\text{1-loop}}\right]_{\text{IR}} = \frac{i}{16\pi^2\epsilon}\mathcal{A}_n^{\text{tree}}\sum_{i\neq j}^n s_{ij}\log\left(s_{ij}\right)+\mathcal{O}\left(\epsilon^0\right).
\end{equation}
We see that the coefficient of the IR divergence is a \textit{transcendental} function. We know, however, that the coefficients of UV divergences are always \textit{rational} functions, since they must be removable by adding local counterterms. It follows that there can never be any UV/IR mixing at one-loop in perturbative quantum gravity and hence that the complete UV divergence is given by the cut-constructible part of the amplitude. This completes the proof of the non-renormalization theorem.

It is important to note that this theorem is valid independent of any anomalies in the duality symmetries. Indeed, in the absence of additional massless degrees of freedom, we expect a non-vanishing ABJ anomaly in the duality currents $j_D^\mu$ \cite{Agullo:2016lkj,Marcus:1985yy}. Explicitly for the $N=1$ case:
\begin{equation}
\langle \nabla_\mu j_D^\mu\rangle = \frac{1}{24\pi^2}R_{\mu\nu\rho\sigma}\tilde{R}^{\mu\nu\rho\sigma}.
\end{equation}
This is a mixed-gravitational anomaly. The question of how this manifests in on-shell scattering amplitudes in the context of $\mathcal{N}=4$ supergravity has been a subject of recent interest \cite{Carrasco:2013ypa,Bern:2017rjw}. Such an anomalous violation of the $U(N)$-invariance at one-loop can appear only in the rational part of the amplitude since the cut-constructible part is completely fixed by unitarity cuts into tree-level amplitudes. The anomaly is therefore irrelevant to the effects of duality invariance on non-renormalization at one-loop. At two-loops however, anomalous rational one-loop amplitudes will have a noticeable effect on ultraviolet divergences and may lead to the renormalization of duality violating six-derivative operators. This question deserves further study. 

\subsection{RG Flow and the Multi-Charge Weak Gravity Conjecture}

With the non-renormalization theorem proven in the previous section, we now show how the argument given in \cite{Charles:2019qqt} generalizes to the multi-charge case. By simple dimensional analysis we know that the counter-terms to one-loop divergences in Einstein-Maxwell are four-derivative operators. In appendix \ref{appA} we give a complete classification of local matrix elements corresponding to four-derivative operators, so together with the non-renormalization theorem proven in the previous section we know that most general \textit{local} UV divergence is given by
\begin{equation}
\label{form}
    \left[\mathcal{A}_4^{\text{1-loop}}\left(1_{\gamma,i}^+,2_{\gamma, j}^+,3_\gamma^{-,k},4_{\gamma}^{-,l}\right)\right]_{\text{UV}} = \frac{c}{16\pi^2\epsilon}\left({\delta_i}^k{\delta_j}^l+{\delta_i}^l{\delta_j}^k\right)[12]^2\langle 34\rangle^2.
\end{equation}

At one-loop, the divergence fixes the dependence of the scattering amplitude on the renormalization group scale $\mu^2$. After adding a counterterm with coefficient $\alpha (\mu) $ to remove the UV divergence, the \textit{physical} scattering amplitude should be independent of $\mu^2$
\begin{equation}
    \mathcal{A}_4^{\text{1-loop}}\left(1_{\gamma,i}^+,2_{\gamma, j}^+,3_\gamma^{-,k},4_{\gamma}^{-,l}\right) = \left[\alpha(\mu^2)+\frac{c}{8\pi^2}\log(\mu^2)\right]\left({\delta_i}^k{\delta_j}^l+{\delta_i}^l{\delta_j}^k\right)[12]^2\langle 34\rangle^2+\mathcal{O}\left(\epsilon^0\right),
\end{equation}
which gives the logarithmic running of the Wilson coefficient
\begin{equation}
    \alpha(\mu^2) = -\frac{c}{8\pi^2}\log\left(\frac{\mu^2}{\Lambda_{\text{UV}}^2}\right),
\end{equation}
where $\Lambda_{\text{UV}}$ is some UV matching scale, assumed to be arbitrarily larger than the horizon scale. The ultraviolet divergence in Einstein-Maxwell coupled to $N$ $U(1)$ gauge fields was first calculated long-ago \cite{Deser:1974cz,Deser:1974xq}, and then recalculated using unitarity methods \cite{Dunbar:1995ed,Norridge:1996he} 
\begin{equation}
    \label{expUV}
    \left[\mathcal{A}_4^{\text{1-loop}}\left(1_{\gamma,i}^+,2_{\gamma, j}^+,3_\gamma^{-,k},4_{\gamma}^{-,l}\right)\right]_{\text{UV}} = \frac{1}{16\pi^2\epsilon}\left(\frac{137    }{120}+\frac{N-1}{20}\right)\left({\delta_i}^k{\delta_j}^l+{\delta_i}^l{\delta_j}^k\right)[12]^2\langle 34\rangle^2.
\end{equation}
This gives the RG coefficient in (\ref{form}) as 
\begin{equation}
    c = \frac{137    }{120}+\frac{N-1}{20}.
\end{equation}
From this matrix element we can reverse engineer the corresponding four-derivative operator
\begin{equation}
    S \supset \alpha(\mu^2)\left(\delta_{ik}\delta_{jl}+\delta_{il}\delta_{jk}\right)\int \text{d}^4 x \sqrt{-g}\left[ \left(F^i_{\mu \nu} F^{j \, \mu \nu} F^k_{\rho \sigma} F^{l \, \rho \sigma} +  F^i_{\mu \nu} \tilde{F}^{j \, \mu \nu} F^k_{\rho \sigma} \tilde{F}^{l \, \rho \sigma}\right)\right].
\end{equation}
Note that we have lost manifest duality invariance when passing from on-shell scattering amplitudes to the effective action and so have made the replacement ${\delta_i}^j\rightarrow \delta_{ij}$. As an important cross-check, the effect of such an operator on the perturbed metric at leading order in $\alpha$ is given by (\ref{metricshift}) to be
\begin{equation}
    \Delta g^{rr} = -\frac{24\alpha(\mu^2)}{15r^6}\sum_{i=1}^N\left(q_i^2+p_i^2\right),
\end{equation}
which manifests the expected electromagnetic duality symmetry, further enhanced to $O(2N)$. 

When evaluating the extremality form, $\mu$ should be taken to be the horizon scale $\mu^2 \sim M_{\text{Pl}}^4/M^2 \sim M_{\text{Pl}}^2/Q^2 $. Since $c>0$, as $Q^2\rightarrow \infty$ the logarithmic term becomes \textit{large} and \textit{positive}. With the logarithmic running included the extremality form at the horizon scale is given by
\begin{align} 
\label{formlog}
    T(q^i,p^i) &= \frac{1}{8\pi^2}\left(\frac{137    }{120}+\frac{N-1}{20}\right)(Q^2)^2\log\left(\frac{\Lambda_{\text{UV}}^2 Q^2}{M_{\text{Pl}}^2}\right)+\alpha^{\text{UV}}_{ijkl} \, (q^i q^j - p^i p^j)(q^k q^l - p^k p^l) \nonumber\\
    &  \hspace{15mm} + \, 8 \beta^{\text{UV}}_{ijkl} q^i p^j q^k p^l - \gamma^{\text{UV}}_{ij} \, \left( q^i q^j - p^i p^j\right)Q^2+ 4 \,  \chi^{\text{UV}}_{ijkl} \, q^i p^j \, \left(q^k q^l-p^k p^l\right)\nonumber\\
    & \hspace{15mm} - 2 \,  \omega^{\text{UV}}_{ij} \,   q^i p^j Q^2,
\end{align}
where $Q^2 =\sum_i(q_i^2+p_i^2)$. In this expression $\alpha^{\text{UV}}$, $\beta^{\text{UV}}$, $\gamma^{\text{UV}}$, $\chi^{\text{UV}}$, and $\omega^{\text{UV}}$ refers to the values of the Wilson coefficients at the matching scale $\Lambda_{\text{UV}}$. Importantly, the logarithmic term is $O(2N)$ invariant and therefore gives an isotropic contribution to the extremality form. This contribution scales like $Q^4 \log Q$, while the rest of the terms scale like $Q^4$. Therefore it dominates over all other contributions. We conclude that for sufficiently large $Q^2$, the extremality form is positive, independent of the values of the Wilson coefficients at the matching scale $\Lambda_{\text{UV}}$, and consequently the multi-charge WGC is always satisfied in the black hole regime.

Here the full $U(N)$ duality invariance of the UV divergence (enhanced to $O(2N)$ in the quartic form) was essential to the argument. It would not have been enough that some Wilson coefficients had a positive logarithmic running, to prove the multi-charge WGC we require positivity in all directions, which as we have shown follows from a generalized non-renormalization theorem as a consequence of tree-level $U(N)$ duality symmetry of Einstein-Maxwell.

It is interesting to note that we can \textit{almost} reach this same conclusion without knowing the explicit form of the UV divergence (\ref{expUV}). In \cite{Hamada:2018dde} the causality bound (\ref{causality}) was applied to the Wilson coefficients at the UV matching scale $\Lambda_{\text{UV}}$ and consequently to constrain the properties of the states integrated out. But this bound must remain valid even deeper in the IR where, as we have seen, the logarithmic running dominates. If the RG coefficient $c$ had been negative, then the bound (\ref{causality}) is eventually violated, indicating the presence of superluminal propagation at very low energies. Since we expect that Einstein-Maxwell is not inconsistent in the deep IR, it must be the case that $c \geq 0$ even without doing a detailed one-loop calculation. This argument has nothing to say about the possibility that $c=0$. Only an explicit calculation is sufficient to demonstrate the existence of a non-vanishing one-loop divergence.

\section{Discussion}

In this paper we have studied the effect of higher-derivative corrections on black hole decay in a more general setting than what has been considered before by allowing for more than one gauge field and by considering in detail the effect of magnetic charges. This conjecture takes a variety of forms, as reviewed in the introduction, but it can be interpreted as the statement that in a UV complete model of quantum gravity there is not an infinite tower of stable states without a symmetry that protects them. Thus, the relevant question is: can nearly extremal charged black holes decay? In particular, we study whether the higher-derivative correction make kinematically possible the process where one large black hole decays into multiple smaller ones.

We find two main differences from the case of a single electric charge: (a) parity-odd operators contribute to the shifted extremality bound when magnetic charges are included (even in the case of a single gauge field), and (b) with multiple charges, allowing extremal black holes to decay imposes a condition on the convex hull of the shifted extremality bound.

The result of our calculation is the shift to the extremality bound for large black holes. The maximum charge in the corrected theory is equal to the mass plus a small correction that is proportional to the coefficients of the higher-derivative operators, seen in (\ref{extremality result}). When there is only one charge, extremal black holes can decay as long as the smaller black holes have a higher charge-to-mass ratio than the large ones. When more charges are present, different generalizations of this condition are possible, but the correct one is a convex hull of the space of allowed charges-to-mass ratios. That paper shows that black holes may decay when the convex hull of the particle spectrum in $\vec{z}$-space contains the unit ball. The condition we have found is that large black holes are able to decay when the convex hull of the allowed charge-to-mass ratios of small black holes contains the unit ball. In our setting, however, this does not play a role; in the regime where we can apply the EFT approach we have outlined, the four-derivative corrections are much smaller than the two derivative terms. In this case, the corrections are always small enough that the space of allowed charge-to-mass ratios is convex anyways. Therefore, we are interested in the simpler requirement that the shift to extremality is always positive. 

There are a number of arguments that attempt to establish the weak gravity conjecture and we have outlined how some of them might apply to multi-charged black holes. In addition to reviewing the arguments from unitarity and causality, we have shown how to extend the argument of \cite{Charles:2019qqt} to the multi-charged case. In doing so, we have presented what we believe is a novel proof of the statement that only duality-invariant terms in the Lagrangian are renormalized. This argument requires that electromagnetic duality is not broken at two-derivative order. It would be interesting to study generalizations where the duality is broken at leading order, such as when a dilaton couples to the field strength. Moreover, this argument depends in an essential way on a symmetry of Einstein-Maxwell which is only present in four-dimensions. In $d\neq 4$ there is no reason to expect that such a non-renormalization theorem should be valid and so it is not clear if the weak gravity conjecture is similarly trivialized by non-trivial RG running.  

Considering scalar fields might also offer the opportunity to check whether the conditions we discuss on Wilson coefficients are satisfied in specific models. One such example is the four-dimensional STU model \cite{Cvetic:1999xp}, which retains four Abelian gauge fields and three dilatonic scalar fields. More generally, the photon and graviton are often accompanied by light scalar moduli in UV complete models from string compactifications. This means that a full understand of the relationship between the weak gravity conjecture and higher-derivative corrections requires studying the role played by scalar fields. Another possibility is to allow for other geometries. Anti-de Sitter space, in particular, presents an interesting opportunity because of the possibility that the AdS/CFT correspondence provides more rigorous bounds on Wilson coefficients (see, for instance \cite{Hofman:2008ar}). We leave these and other generalizations to future work.


\section*{Acknowledgements} 


We would like to thank James T. Liu for collaborating during the early stages of this work. Additionally, we would like to thank Anthony Charles, Henriette Elvang, Finn Larsen and Gary Shiu for useful discussions, and Brando Bellazinni, Javi Serra, and Matt Lewandowski for useful comments after the first version of this paper was submitted. This work was supported in part by the US 
Department of Energy under Grant No.~DE-SC0007859. CRTJ was supported in part by a 
Leinweber Student Fellowship and in part by a Rackham Predoctoral 
Fellowship from the University of Michigan.


\appendix


\section{EFT Basis and On-Shell Matrix Elements}
\label{appA}

Operator redundancies in EFTs arise due to the field reparametrization invariance of physical observables \cite{Arzt:1993gz}. For example, in Einstein-Maxwell we consider redefinitions of the metric of the form
\begin{equation}
g'_{\mu\nu} \equiv g_{\mu\nu} + c_1R_{\mu\nu} + c_2R g_{\mu\nu}+c_3 F_{\mu\rho}{F^\rho}_\nu+...
\end{equation}
where $c_i$ are independent coefficients. In the complete effective action (including all possible terms of all mass dimensions consistent with the assumed symmetries) the effect of such a field redefinition is to shift the Wilson coefficients. By choosing $c_i$ in a particular way, certain operators can be removed from the effective action entirely; these are the so-called redundant operators. One approach to constructing a non-redundant basis of operators is to first enumerate all local operators, then use the most general field reparametrization to remove redundant operators. In this appendix we describe an alternative approach that makes use of on-shell scattering amplitudes methods. 

The S-matrix corresponding to the effective action is likewise a physical observable, and independent of the choice of field parametrization. In the tree approximation, gauge invariant effective operators generate Lorentz invariant on-shell matrix elements without kinematic singularities. The on-shell method begins with the observation that there is a one-to-one correspondence between non-redundant gauge invariant local operators and Lorentz invariant local matrix elements \cite{Henning:2017fpj}. By making use of the spinor-helicity formalism for massless on-shell states \cite{Elvang:2015rqa}, it is sometimes more efficient to construct an independent set of the latter. Below we use this correspondence to construct a complete basis for operators coupling gravity to $N$ $U(1)$ gauge fields with up to four derivatives.

The on-shell matrix elements we construct are in the helicity basis. Lorentz invariance is encoded in the requirement that the expressions we construct are rational functions of spinor brackets
\begin{equation}
	\langle ij \rangle = \epsilon^{\dot{\alpha}\dot{\beta}} \tilde{\lambda}_{i\dot{\alpha}} \tilde{\lambda}_{j\dot{\beta}}, \hspace{5mm} [ ij ] = \epsilon_{\alpha\beta} \lambda_i^\alpha \lambda_j^\beta.
\end{equation}
On-shell matrix elements corresponding to gauge invariant local operators are given by polynomials of spinor brackets; we first construct a basis of monomials satisfying certain physical conditions. The first condition we impose is consistency with the action of the massless little group. Such monomials must scale homogeneously with the correct little group weight determined by the helicities $h_i$ of each of the external states
\begin{equation}
M\left(t\lambda_i, t^{-1}\tilde{\lambda}_i\right) = t^{2h_i}M\left(\lambda_i, \tilde{\lambda}_i\right) \, .
\end{equation}
Here we are scaling the spinors of particle $i$ separately, leaving the remaining spinors unchanged. Since the expressions we are constructing are simply strings of $\tilde{\lambda}$s and $\lambda$s, this constraint is equivalent to the following 
\begin{equation}
  2 h_i = (\text{\# of }\lambda_i)-(\text{\# of }\tilde{\lambda}_i).
\end{equation}
This constraint places a lower bound on the mass dimension of the monomial. The minimal dimension monomial we could construct with the correct little group weight for each state contains no anti-holomorphic spinors ($\tilde{\lambda}$) for positive helicity states, no holomorphic spinors ($\lambda$) for negative helicity states and no spinors of either chirality for helicity zero states. As an example, the schematic form of such a minimal dimension monomial
\begin{equation}
  M_4\left(1^{+2},2^{+1},3^{-2},4^0\right) \sim \lambda_1^4 \lambda_2^2 \tilde{\lambda}_3^4. 
\end{equation}
As described above, we need to contract the implicit spinor indices in all inequivalent ways to form a basis of such monomials. The mass dimension of such a string is given simply by $[\lambda]=[\tilde{\lambda}]=1/2$. In this example the minimal dimension is 5. Non-minimal monomials may be generated by introducing further pairs of spinors $\lambda_i \tilde{\lambda}_i \sim p_i$, which have zero little group weight. In general, for a monomial with $k$ photon states and $m$ graviton states the dimension of the monomial is bounded below as:
\begin{equation}
  [M_n] \geq k + 2m.
\end{equation}
To connect this to the EFT basis, such a monomial must correspond to the Feynman vertex rule derived from a gauge invariant local operator. Since polarization vectors for Bosonic states are dimensionless, $[\epsilon]=0$, the mass dimension of the monomial can only arise from powers of momenta generated from derivative interactions. For a local operator with $D$ derivatives the matrix element of $k$ photons and $m$ gravitons has the schematic form
\begin{equation} \label{mnfeyn}
	M_n\left(\{\epsilon,p\}\right) \sim  \epsilon_\gamma^k \epsilon_h^m p^D,
\end{equation}
and so the dimension of the monomial is simply 
\begin{equation}
  [M_n]=D.
\end{equation}
Putting these results together we find that the number of photons and gravitons in a local matrix element is bounded above by the number of derivatives in the corresponding local operator
\begin{equation} \label{Dbound}
D \geq k + 2m.
\end{equation}
This also bounds the total number of states $n=k+m$ (since both $k$ and $m$ are non-negative) as $D\geq n$. Our task is now to enumerate all inequivalent monomials for photon and gravitons with $D=3$ and $D=4$ and identify the corresponding local operators. Here inequivalent means constructing a basis of monomials that are not related to each other by momentum conservation
\begin{equation}
  \sum_{j=1}^n \langle i j\rangle [jk] = 0, 
\end{equation}
 or Schouten identities
\begin{equation}
  \langle ij \rangle \langle kl\rangle + \langle ik\rangle \langle lj\rangle + \langle il\rangle \langle jk\rangle = 0, \hspace{10mm} [ij][kl]+[ik][lj]+[il][jk]=0. 
\end{equation}
A straightforward (though certainly not optimal) approach to this is to first generate a complete basis of monomials, and then numerically evaluate on sets of randomly generated spinors to find a linearly independent subset.

To construct local operators corresponding to the monomials we can make use of the following replacement rules, for photons:
\begin{align}
  \lambda_\alpha\lambda_\beta \rightarrow F^+_{\alpha\beta}\equiv \sigma^{\mu\nu}_{\alpha\beta}F_{\mu\nu}, \hspace{10mm} \tilde{\lambda}_{\dot{\alpha}}\tilde{\lambda}_{\dot{\beta}} \rightarrow F^-_{\dot{\alpha}\dot{\beta}} \equiv \overline{\sigma}^{\mu\nu}_{\dot{\alpha}\dot{\beta}}F_{\mu\nu},
\end{align}
and for gravitons\footnote{Here we are defining $\sigma^{\mu\nu}_{\alpha\beta} \equiv \frac{i}{4}\epsilon^{\dot{\alpha}\dot{\beta}}
\left(\sigma^\mu_{\alpha\dot{\alpha}} \sigma^\nu_{\beta\dot{\beta}}-\sigma^\nu_{\alpha\dot{\alpha}} \sigma^\mu_{\beta\dot{\beta}}\right)$ and $\overline{\sigma}^{\mu\nu}_{\dot{\alpha}\dot{\beta}} \equiv \frac{i}{4}\epsilon^{\alpha\beta}
\left(\sigma^\mu_{\alpha\dot{\alpha}} \sigma^\nu_{\beta\dot{\beta}}-\sigma^\nu_{\alpha\dot{\alpha}} \sigma^\mu_{\beta\dot{\beta}}\right)$. Using standard trace identities, we can rewrite the local operators we construct in the more familiar (though less compact) Lorentz vector notation.}:
\begin{align}
  \lambda_\alpha\lambda_\beta\lambda_\gamma\lambda_\delta \rightarrow W^+_{\alpha\beta\gamma\delta} \equiv \sigma^{\mu\nu}_{\alpha\beta}\sigma^{\rho\sigma}_{\gamma\delta}W_{\mu\nu\rho\sigma}, \hspace{5mm} \tilde{\lambda}_{\dot{\alpha}}\tilde{\lambda}_{\dot{\beta}}\tilde{\lambda}_{\dot{\gamma}}\tilde{\lambda}_{\dot{\delta}} \rightarrow  W^-_{\dot{\alpha}\dot{\beta}\dot{\gamma}\dot{\delta}} \equiv\overline{\sigma}^{\mu\nu}_{\dot{\alpha}\dot{\beta}}\overline{\sigma}^{\rho\sigma}_{\dot{\gamma}\dot{\delta}}W_{\mu\nu\rho\sigma},
\end{align}
where $F^\pm$ and $W^\pm$ are the (anti-)self-dual field strength and Weyl tensors respectively. For non-minimal operators there are additional helicity spinors; these must come in pairs with zero net little group weight and so we can replace:
\begin{equation}
  \lambda^i_\alpha \tilde{\lambda}^i_{\dot{\alpha}} \rightarrow \sigma^\mu_{\alpha\dot{\alpha}}\nabla_\mu \, ,
\end{equation}
where the derivative acts on the local operator creating state $i$. As an illustrative example, consider the following matrix element
\begin{align} \label{matrixex}
  &M_4\left(1^{+1},2^{+1},3^{-1},4^{-2}\right)\nonumber\\
  &= [12]^3\langle 34 \rangle^2 \langle 14 \rangle \langle 24 \rangle \nonumber\\
  &= (\lambda_{1}^{\alpha_1} \lambda_{1}^{\alpha_2})(\lambda_{2\alpha_1}\lambda_{2\alpha_2})(\tilde{\lambda}_{3\dot{\alpha}_1}\tilde{\lambda}_{3\dot{\alpha}_2})(\tilde{\lambda}_{4}^{\dot{\alpha}_1}\tilde{\lambda}_{4}^{\dot{\alpha}_2}\tilde{\lambda}_{4}^{\dot{\alpha}_3}\tilde{\lambda}_{4}^{\dot{\alpha}_4})(\tilde{\lambda}_{1\dot{\alpha}_3}\lambda_1^{\alpha_3})(\tilde{\lambda}_{2\dot{\alpha}_4}\lambda_{2\alpha_3}) \, .
\end{align}
Using the replacement rules given above, this can be generated from the following local operator
\begin{equation}
  [12]^3\langle 34 \rangle^2 \langle 14 \rangle \langle 24 \rangle  \rightarrow \epsilon^{\dot{\alpha}_3\dot{\alpha}_4}\sigma^\mu_{\alpha_3\dot{\alpha}_3}\sigma^\nu_{\alpha_4\dot{\alpha}_4}(\nabla_\mu F^{1+\alpha_1 \alpha_2})(\nabla_\nu F^{2+}_{\alpha_1\alpha_2})F^{3-}_{\dot{\alpha}_1\dot{\alpha}_2}W^{-\dot{\alpha}_1\dot{\alpha}_2\dot{\alpha}_3\dot{\alpha}_4}
\end{equation}
Here we have used a superscript $F^{i}$ to indicate that the spin-1 states correspond to distinct $U(1)$ gauge groups. If two or more states with the same helicity correspond to the same $U(1)$ factor, then we must Bose symmetrize over the particle labels in the matrix elements before applying the replacement rules. This generically reduces the number of independent local operators at a given order in the derivative expansion. 

Finally we must discuss the constraints of parity conservation. In the spinor-helicity formalism, parity $\mathcal{P}$ acts by interchanging the chirality of the spinors $\lambda_{i\alpha}\leftrightarrow \tilde{\lambda}_{i\dot{\alpha}}$, or equivalently interchanging angle and square spinor brackets\footnote{This definition of parity makes sense only if we write the entire matrix element in terms of spinor brackets. For example, to see that local matrix elements containing a single instance of the Levi-Civita symbol are parity odd we must use the identity $\epsilon^{\mu\nu\rho\sigma}p_{1\mu}p_{2\nu}p_{3\rho}p_{4\sigma} \propto [12]\langle 23 \rangle [34] \langle 41\rangle - \langle 12 \rangle [23] \langle 34 \rangle [41]$. }. A local operator is called parity conserving if it generates local matrix elements that satisfy
\begin{equation}
 \mathcal{P}\cdot M_n\left(1^{h_1},2^{h_2},..., n^{h_n}\right) = M_n\left(1^{-h_1},2^{-h_2},..., n^{-h_n}\right).
\end{equation}
This means that when constructing a basis of local operators using the method described above, in a parity conserving model the matrix elements $M_n\left(1^{h_1},2^{h_2},..., n^{h_n}\right)$ and $M_n\left(1^{-h_1},2^{-h_2},..., n^{-h_n}\right)$ should not be counted separately, while in a parity non-conserving model they should be. \subsection{Three-Derivative Operators}
In accord with the constraint (\ref{Dbound}) the possible, non-redundant, three-derivative operators that generate on-shell matrix elements with $k$-photons and $m$-gravitons have
\begin{equation}
  (k,m) \in \{(3,0)\}.
\end{equation}
The list of possible matrix elements modulo Schouten and momentum conservation, and the corresponding local operators is:\\
$(+1,+1,+1):$
\begin{align}
  [12][23][31] \rightarrow F^{1+}_{\alpha\beta}F^{2+\beta\gamma}{{F^{3+}}_\gamma}^\alpha.
\end{align}
$(-1,-1,-1):$
\begin{align}
  \langle 12 \rangle \langle 23\rangle \langle 31\rangle \rightarrow  F^{1-}_{\dot{\alpha}\dot{\beta}}F^{2-\dot{\beta}\dot{\gamma}}{{F^{3-}}_{\dot{\gamma}}}^{\dot{\alpha}}.
\end{align}
There are two independent, three-derivative local operators. Imposing parity conservation there is only a single independent local operator. Such operators vanish unless all field strength tensors are from distinct $U(1)$ factors. To preserve Bose symmetry of the matrix element we see that the associated Wilson coefficients must be totally antisymmetric in flavor indices.

An equivalent form of the three-derivative effective Lagrangian is
\begin{equation}
  \mathcal{L}^{(3)} = a_{ijk}F^{i}_{\mu\nu}F^{j\nu\rho}{F^{k}_\rho}^\mu + b_{ijk}F^{i}_{\mu\nu}F^{j\nu\rho} \tilde{F}_\rho^{k\mu},
\end{equation}
where both $a_{ijk}$ and $b_{ijk}$ are totally antisymmetric. The first operator ($a$) is parity even while the second ($b$) is parity odd. 
\subsection{Four-Derivative Operators}
The possible, non-redundant, four-derivative operators generate on-shell matrix elements with $k$-photons and $m$-gravitons with
\begin{equation}
  (k,m) \in \{(2,1),(4,0)\}.
\end{equation}
The list of possible matrix elements modulo Schouten and momentum conservation, and the corresponding local operators is :\\
$(+1,+1,+2):$
\begin{align}
  [13]^2[23]^2 \rightarrow F^{1+}_{\alpha_1\alpha_2}F^{2+}_{\alpha_3\alpha_4}W^{+\alpha_1\alpha_2\alpha_3\alpha_4}.
\end{align}
$(-1,-1,-2):$
\begin{align}
  \langle 13\rangle^2\langle 23\rangle^2 \rightarrow F^{1-}_{\dot{\alpha}_1\dot{\alpha}_2}F^{2-}_{\dot{\alpha}_3\dot{\alpha}_4}W^{-\dot{\alpha}_1\dot{\alpha}_2\dot{\alpha}_3\dot{\alpha}_4}.
\end{align}
$(+1,+1,+1,+1):$
\begin{align}
  [13]^2[24]^2 &\rightarrow F^{1+}_{\alpha_1\alpha_2}F^{3+\alpha_1\alpha_2}F^{2+}_{\alpha_3\alpha_4}F^{4+\alpha_3\alpha_4}\nonumber\\
  [12][23][34][41] &\rightarrow F^{1+}_{\alpha_1\alpha_2}F^{2+\alpha_2\alpha_3}F^{3+}_{\alpha_3\alpha_4}F^{4+\alpha_4\alpha_1}\nonumber\\
  [12]^2[34]^2 &\rightarrow F^{1+}_{\alpha_1\alpha_2}F^{2+\alpha_1\alpha_2}F^{3+}_{\alpha_3\alpha_4}F^{4+\alpha_3\alpha_4}.
\end{align}
$(-1,-1,-1,-1):$
\begin{align}
  \langle 13\rangle^2\langle 24\rangle^2 &\rightarrow F^{1-}_{\dot{\alpha}_1\dot{\alpha}_2}F^{3-\dot{\alpha}_1\dot{\alpha}_2}F^{2-}_{\dot{\alpha}_3\dot{\alpha}_4}F^{4-\dot{\alpha}_3\dot{\alpha}_4}\nonumber\\
  \langle 12\rangle \langle 23\rangle \langle 34\rangle \langle 41\rangle &\rightarrow F^{1-}_{\dot{\alpha}_1\dot{\alpha}_2}F^{2-\dot{\alpha}_2\dot{\alpha}_3}F^{3-}_{\dot{\alpha}_3\dot{\alpha}_4}F^{4-\dot{\alpha}_4\dot{\alpha}_1}\nonumber\\
  \langle 12\rangle^2\langle 34\rangle^2 &\rightarrow F^{1-}_{\dot{\alpha}_1\dot{\alpha}_2}F^{2-\dot{\alpha}_1\dot{\alpha}_2}F^{3-}_{\dot{\alpha}_3\dot{\alpha}_4}F^{4-\dot{\alpha}_3\dot{\alpha}_4}.
\end{align}
$(+1,+1,-1,-1):$
\begin{align}
  [12]^2\langle 34 \rangle^2 \rightarrow F^{1+}_{\alpha_1\alpha_2}F^{2+\alpha_1\alpha_2}F^{3-}_{\dot{\alpha}_1\dot{\alpha}_2}F^{4-\dot{\alpha}_1\dot{\alpha}_2}.
\end{align}
There are five independent, four-derivative local operators. Imposing parity conservation there are only three independent local operators. An equivalent form of the four-derivative effective Lagrangian is
\begin{align}
  \mathcal{L}^{(4)} &= \alpha_{ijkl}F^{i}_{\mu\nu}F^{j\mu\nu}F^{k}_{\rho\sigma}F^{l\rho\sigma} + \beta_{ijkl}F^{i}_{\mu\nu}\tilde{F}^{j\mu\nu}F^{k}_{\rho\sigma}\tilde{F}^{l\rho\sigma} + \gamma_{ij}F^{i}_{\mu\nu}F^{j}_{\rho\sigma}W^{\mu\nu\rho\sigma} \nonumber\\
                    &\hspace{10mm}+ \chi_{ijkl}F^{i}_{\mu\nu}F^{j\mu\nu}F^{k}_{\rho\sigma}\tilde{F}^{l\rho\sigma} + \omega_{ij}F^{i}_{\mu\nu}\tilde{F}^{j}_{\rho\sigma}W^{\mu\nu\rho\sigma}.
\end{align}
The first three operators ($\alpha$, $\beta$ and $\gamma$) are parity even, while the remaining two ($\chi$ and $\omega$) are parity odd. 

\section{Corrections to the Maxwell equation}
    \label{appMax}
    
In this appendix we shall review the derivation of (\ref{Maxwell Corrections}). Recall the corrected equation of motion for the gauge field:
\begin{align}
    \begin{split}
    \nabla_{\mu} F^{i \mu \nu} =& \, \nabla_{\mu} \Big( 8 \, \alpha_{ijkl}  F^{j \mu \nu} F^k_{\alpha \beta} F^{l \alpha \beta} + 8 \, \beta_{ijkl}  \tilde{F}^{j \mu \nu} F^k_{\alpha \beta} \tilde{F}^{l \alpha \beta} + 4 \, \gamma_{ij}  F^j_{\alpha \beta} W^{\mu \nu \alpha \beta}  \\
    &  \qquad \qquad + 4 \, \left( \chi_{ijkl} \tilde{F}^{j \mu \nu} F^k_{\alpha \beta} F^{l \alpha \beta} + \chi_{klij} F^{j \mu \nu} \tilde{F}^k_{\alpha \beta} F^{l \alpha \beta} \right) +  4 \, \omega_{ij}  \tilde{F}^j_{\alpha \beta} W^{\mu \nu \alpha \beta} \Big) \, .
    \end{split}
\end{align}
For simplicity we label the term in the parentheses on the right-hand side of (\ref{Maxwell}) by $G^{i \, \mu \nu}$. First note that the anti-symmetry of $F^{\mu \nu}$ allows us to rewrite the equation of motion as
\begin{align}
    \begin{split}
    \frac{1}{\sqrt{-g}} \partial_{\mu} \left[ \sqrt{-g} \, F^{i \mu \nu} \right] =     \frac{1}{\sqrt{-g}}  \partial_{\mu} \left[ \sqrt{-g} \, G^{i \, \mu \nu} \right] \, .
    \end{split}
\end{align}
We expand this equation in power of the coefficients $\alpha, \ ... \ \omega$. The zeroth- and first-order equations are:
\begin{subequations}
\begin{align}
    & \partial_{\mu} \left[ \sqrt{-g} \, F{}^{i \mu \nu} \right]^{(0)} = 0 \label{EOM0}
    \\
    & \partial_{\mu} \left[ \sqrt{-g} \, F{}^{i \mu \nu} \right]^{(1)} = \partial_{\mu} \left[ \sqrt{-
    g} \, G^{i \, \mu \nu}  \right]^{(1)} \, . \label{EOM1}
\end{align}
\end{subequations}
The solution to the zeroth-order equation is the uncorrected Reissner-Nordstr{\"o}m solution. We are interested in obtaining the first-order part, which represents the corrections to the background. The derivative may be removed from (\ref{EOM1}) because an additive constant has the same fall-off in $r$ as the solution to (\ref{EOM0}), so we may absorb it into the definition of integration constant in the zeroth-order solution, which is $q$. As a result, we have 
\begin{align}
    \begin{split}
        \left[ \sqrt{-g} \, F{}^{i \mu \nu} \right]^{(1)} = \left[ \sqrt{-
    g} \, G^{i \, \mu \nu}  \right]^{(1)} \, .
    \end{split}
\end{align}
 Note that $G^{\mu \nu}$ depends explicitly on $(\, \alpha, ..., \omega \, ) $, so $(G^{\mu \nu})^{(1)}$, which is first-order in the coefficients, depends only on the zeroth-order value of the fields $F^{\mu \nu}$ and $W^{\mu \nu \rho \sigma}$. 
 
 In addition to the Maxwell equation, the gauge fields must satisfy the Bianchi identity
\begin{equation}
    \partial_\mu F^{i}_{\nu\rho}+\partial_\nu F^{i}_{\rho\mu}+\partial_\rho F^{i}_{\mu\nu}=0.
\end{equation}
Together with the assumed spherically symmetry, which imposes that only $F^{i}_{t r}$ and $F^{i}_{\theta \phi}$ are non-zero, this gives the following constraint on the \textit{magnetic} component of the gauge field
\begin{equation} \label{magBI}
    \partial_{r}F^{i}_{\theta \phi}=0.
\end{equation}
Since the leading order magnetic field (\ref{zerosol}) is the unique spherically symmetric field with magnetic monopole moment $p^i$, and by (\ref{magBI}) there can be no subleading $1/r$ corrections, it remains the exact solution even with the addition of higher-derivative interactions. Therefore we are only interested in the corrections to the electric fields $F^{(i)}_{t r}$. Using that $g^0_{tt} = - g^0_{rr}$, we have
\begin{align}
    \begin{split}
    & \left[ \sqrt{-g} F{}^{i \, t r} \right]^{(1)} \ = \  \sqrt{-g}^{(0)} \left(  8 \alpha_{ijkl}  F^{(0) j}{}_{t r} F^{(0) k}{}_{tr} F^{(0)l}{}_{tr} + ... \right).
    \end{split}
\end{align}
Now we may use this to compute the first contribution to the stress tensor corrections. This relies on the non-trivial fact that this combination of $\sqrt{-g}$ and $F$ is the only combination that appears in the corrections to the stress tensor. To see this consider the stress tensor for a Maxwell field, 
\begin{align}
\begin{split}
    T_{\mu \nu} = F^i_{\mu \alpha} F^i_{\nu}{}^{\alpha} - \frac{1}{4} F^i_{\alpha \beta} F^{i \alpha \beta} g_{\mu \nu}.
\end{split}
\end{align}
We are interested only in the corrections to 
\begin{align}
\begin{split}
    T_t {}^t = F^i{}_{t \alpha} F^{i t \alpha} - \frac{1}{4} F^i{}_{\alpha \beta} F^i{}^{\alpha \beta} \delta_t{}^t \, .
\end{split}
\end{align}
We use the fact that only $F_{t r}$ and $F_{\theta \phi}$ are non-zero, and only the former is corrected, to write 
\begin{align}
\begin{split}
    T_t{}^t \ =& \ \frac{1}{2} F^i{}_{t r} F^{i t r} - \frac{1}{2} F^i{}_{\theta \phi} F^{i \theta \phi} \\
=& \ (T^{(0)})_t{}^t - \left[  \sqrt{-g} F^{i t r} \right]^{(1)} \left[  \sqrt{-g} F^{i t r} \right]^{(0)} / (g_{\theta \theta} g_{\phi \phi}) + \mathcal{O} \left[ (\alpha, ...)^2 \right] \, .  
\end{split}
\end{align}
So we have found that 
\begin{align}
\begin{split}
    (T^{(1)}_{Max})_t{}^t =& - \left[  \sqrt{-g} F^{i t r} \right]^{(1)} \left[  \sqrt{-g} F^{i t r}     \right]^{(0)} / (g_{ \theta \theta } g_{\phi \phi})  \\
    = & -  \sqrt{-g}^{(0)} \left(  8 \alpha_{ijkl}  F^{(0) j}{}_{t r} F^{(0) k}{}_{t r} F^{(0)l}{}_{t r}  + ... \right) \sqrt{-g}^{(0)} F^{i t r}{}^{(0)} / (g_{\theta \theta } g_{\phi \phi}) \\
    = & \left(  8 \alpha_{ijkl}  F^{(0) j}{}_{t r} F^{(0) k}{}_{t r} F^{(0)l}{}_{t r}  + ... \right)  F^i{}_{t r}{}^{(0)} \, .
\end{split}
\end{align}
Evaluating this expression gives the result obtained in (\ref{Maxwell Corrections}).

\section{Variations of Four-Derivative Operators with respect to the Metric}
    \label{appLag}

In section II, we computed the shift to the geometry by first computing the shift to the stress tensor due to the presence of higher-derivative operators. One source of stress tensor corrections comes from varying the four-derivative operators with respect to the metric. The variations of each of these terms are recorded here for reference.
\begin{align}
    \begin{split}
        (F^i F^j) (F^k F^l): \qquad \qquad & g_{\alpha \beta} (F^i F^j) (F^k \cdot F^l) - 4 \left( F^i_{\mu \alpha} F^{j \mu}{}_{\beta} (F^k  F^l) + (F^i  F^j) F^k_{\mu \alpha} F^{l \mu}{}_{\beta}  \right)  \\[5pt]
        (F^i  \tilde{F}^j) (F^k  \tilde{F}^l) : \qquad \qquad & - g_{\alpha \beta} (F^i  \tilde{F}^j) (F^k  \tilde{F}^l) \\[5pt]
        W F^i F^j :  \qquad \qquad & g_{\alpha \beta} W F^i F^j - 3 R^{\mu}{}_{\alpha \rho \sigma} (F^i_{\mu \beta} F^{j \rho \sigma}+ F^{i \rho \sigma} F^j_{\mu \beta} ) + 4 R_{\alpha \mu} (F^i_{\beta \nu} F^{j \mu \nu} + F^{i \mu \nu} F^j_{\beta \nu} )\\
            & \qquad + 4 R_{\mu \nu} F^{i \mu}{}_{\alpha} F^{j \nu}{}_{\beta}  - \frac{4}{3} R F^i_{\alpha \mu} F^j_{\beta}{}^{\mu} - \frac{2}{3} R_{\alpha \beta} (F^i  F^j) \\
            & \qquad - 4 \nabla_{\mu} \nabla_{\nu} (F^{i \mu}{}_{\alpha}  F^{j \nu}{}_{\beta}) - 4 \nabla_{\mu} \nabla_{\alpha} (F^{i \mu}{}_{\nu} F^j_{\beta}{}^{\nu}) + 2 g_{\alpha \beta} \nabla_{\mu} \nabla_{\nu} (F^{i \mu}{}_{\rho}  F^{j \nu \rho}) \\
            & \qquad + 2 \Box (F^i_{\alpha \mu} F^j_{\beta}{}^{\mu}) + \frac{2}{3} \nabla_{\alpha} \nabla_{\beta} (F^i  F^j) - \frac{2}{3} g_{\alpha \beta} \Box (F^i  F^j) \\[5pt]
        (F^i  \tilde{F}^j) (F^k  F^l) :  \qquad \qquad & -4 (F^i  \tilde{F}^j) F^k_{\mu \alpha} F^{l \mu}{}_{\beta} \\[5pt]
        W F^i \tilde{F}^j:   \qquad \qquad & - 2 R^{\mu}{}_{\alpha \rho \sigma} F^i_{\mu \beta} \tilde{F}^{j \rho \sigma} + 4 R_{\alpha \mu} F^i_{\beta \nu} \tilde{F}^{j \mu \nu}  - \frac{2}{3} R_{\alpha \beta} (F^i  \tilde{F}^j) \\
            & \qquad - 4 \nabla_{\mu} \nabla_{\nu} (F^{i \mu}{}_{\alpha}  \tilde{F}^{j \nu}{}_{\beta}) - 4 \nabla_{\mu} \nabla_{\alpha} (F^{i \mu}{}_{\nu} \tilde{F}^j_{\beta}{}^{\nu}) + 2 g_{\alpha \beta} \nabla_{\mu} \nabla_{\nu} (F^{i \mu}{}_{\rho}  \tilde{F}^{j \nu \rho} )\\
            & \qquad + 2 \Box (F^i_{\alpha \mu} \tilde{F}^j_{\beta}{}^{\mu}) + \frac{2}{3} \nabla_{\alpha} \nabla_{\beta} (F^i  \tilde{F}^j) - \frac{2}{3} g_{\alpha \beta} \Box (F^i  \tilde{F}^j)
    \end{split}
\end{align}
Each of the terms on the left-hand side are multiplied by $\sqrt{-g}$ in the action. Note that we use the shorthand $( F^i  F^j) $ to denote $F^i_{\mu \nu} F^{j \mu \nu}$, and $W A B$ to denote $W_{\mu \nu \rho \sigma} A^{\mu \nu} B^{\rho \sigma}$.

\section{Proof of Convexity of the Extremality Surface}
    \label{appConvex}
    
In this appendix we give a short proof of the claim made in section \ref{sec:decay}, that in the perturbative regime, $Q^2\gg 1$, the extremality surface bounds a convex region. Though convexity is a global property, we can reduce the problem to a local one through the \textit{Tietze-Nakajima theorem} \cite{bjorndahl_karshon_2010}: if $X\subset \mathds{R}^n$ is closed, connected and \textit{locally convex}, then $X$ is convex. Here local convexity means that for each $x\in X$, for some $\delta>0$ the set $B_\delta(x)\cap X$ is convex. 
    
Since the requirements of closure and connectedness are trivial for the kinds of regions we are considering, it remains to show that the extremality surface is the boundary of a locally convex set. The key idea of the argument is to show that on a sufficiently small neighborhood of any point, the surface is well approximated by an inverted paraboloid up to $\mathcal{O}(1/Q^2)$ corrections. Local convexity is then a consequence of the convexity of the paraboloid hypograph.
    
Consider a general co-dimension-1 hypersurface $X$ embedded in $\mathds{R}^n$, defined by an equation of the form
    \begin{equation}
        \label{defS}
        \sum_{i=1}^n x_i^2 = 1 + T(x_i),
    \end{equation}
    where $T(x_i)$ is \textit{small} in the sense that 
    \begin{equation}
        \label{smallness}
        \biggr|\sum_{i=1}^n x_i^2-1\biggr| < \epsilon, 
    \end{equation}
for all points $x_i\in X$, for some arbitrarily small $\epsilon >0$. Since this condition is preserved under orthogonal rotations, every point on $X$ can be mapped to $x_i=0$ for $i>1$ up to a redefinition of the function $T(x_i)$. Without loss of generality then we will study the local neighbourhood of such a point. We use the fact that we are interested in functions of the form
    \begin{equation}
        T(x_i) = \sum_{ijkl}T_{ijkl}x_ix_jx_kx_l \, .
    \end{equation}
Here the smallness condition (\ref{smallness}) is equivalent to the statement that $|T_{ijkl}|\sim \epsilon$. To begin with we can rewrite the equation (\ref{defS}) in a useful form
    \begin{align}
        x_1^2 = 1-\sum_{i\neq 1} x_i^2 &+T_{1111}x_1^4+ 4x_1^3\sum_{i}T_{111i}x_i+ 6x_1^2 \sum_{ij\neq 1}T_{11ij}x_i x_j \nonumber\\ 
        &+4x_1\sum_{ijk\neq 1}T_{1ijk}x_i x_j x_k   +\sum_{ijkl\neq 1}T_{ijkl}x_i x_j x_k x_l.
    \end{align}
At $x_i=0$, $i>1$, for small $\epsilon$ there is a single value of $x_1>0$ on $X$. Since we are interested in the surface on an arbitrarily small convex neighbourhood $D$ of $x_i=0$, $i>1$, we can construct a local parametrization of the surface as a function $x_1:D\rightarrow \mathds{R}$
    \begin{equation}
    \label{local}
        x_1(x_2,...,x_n) = 1-\frac{1}{2}\sum_{i\neq 1}x_i^2+\frac{1}{2}T_{1111} +\frac{1}{2}T_{1111}\sum_{i\neq 1}x_i^2+ 3\sum_{i}T_{111i}x_i+ 3 \sum_{i,j\neq 1}T_{11ij}x_i x_j + \mathcal{O}(x_i^3).
    \end{equation}
It is an elementary theorem that the \textit{hypograph} of a function $f:D\rightarrow \mathds{R}$, with $D$ a convex set in $\mathds{R}^{n-1}$, is a convex set in $\mathds{R}^n$ if the Hessian of $f$ is negative definite on the interior of $D$. From (\ref{local}) we can read off the eigenvalues of the Hessian matrix at this point as $-1 + \mathcal{O}(\epsilon)$. Since the eigenvalues of the Hessian are continuous on $X$ they must all be negative on some neighbourhood of this point. This completes the proof that $X$ is locally convex.
    
\bibliographystyle{JHEP}
\bibliography{cite.bib}
\end{document}